%% file: schlieren1.tex
\documentclass[preprint,showpacs,preprintnumbers,amsmath,amssymb,superscriptaddress]{revtex4}

\usepackage{graphicx}
\usepackage{dcolumn}
\usepackage{bm,bbm}
\usepackage{color}
\usepackage{placeins}

\usepackage{hyperref}
\hypersetup{
    colorlinks,
    citecolor=black,
    filecolor=black,
    linkcolor=black,
    urlcolor=black
}

\input{mathcommands.tex}

\begin{document}


\title{Machine learning methods for Schlieren imaging of a plasma channel in tenuous atomic vapor}%

\author{G\'{a}bor B\'{\i}r\'{o}}

\author{Mih\'{a}ly Pocsai}

\author{Imre F. Barna}
\affiliation{Wigner Research Centre for Physics, Budapest, Hungary}

\author{Joshua T. Moody}
\affiliation{Max Planck Institute for Physics, Munich, Germany}

\author{G\'{a}bor Demeter}
\affiliation{Wigner Research Centre for Physics, Budapest, Hungary}



\date{\today}

\begin{abstract}
We investigate the usage of a Schlieren imaging setup to measure the geometrical dimensions of a plasma channel in atomic vapor. Near resonant probe light is used to image the plasma channel in a tenuous vapor and machine learning techniques are tested for extracting quantitative information from the images. By building a database of simulated signals with a range of plasma parameters for training  Deep Neural Networks, we demonstrate that they can extract from the Schlieren images reliably and with high accuracy the location, the radius and the maximum ionization fraction of the plasma channel as well as the width of the transition region between the core of the plasma channel and the unionized vapor. We test several different neural network architectures with supervised learning and show that the parameter estimations supplied by the networks are resilient with respect to slight changes of the experimental parameters that may occur in the course of a measurement.
\end{abstract}

\pacs{}
\maketitle

\section{Introduction}

Rapid developments in computing and information science in recent years led to  increasingly sophisticated Machine Learning (ML) implementations. The list of possible applications is ever growing, including (but not limited to) autonomous driving~\cite{NARAYANAN2020255} , healthcare~\cite{VERMA20222144}, speech recognition~\cite{6423821} and various high-energy physics studies \cite{Feickert:2021ajf, Biro:2021zgm}. Machine learning methods have been used for some time also for evaluating optical diagnostic measurements in plasma physics, for example, tomographic measurements of radiation from fusion plasmas \cite{Matos2017, Mlynar2019, Clayton2013, Demeter1997}.

Schlieren imaging is a sensitive method for the detection of refractive index variations in transparent media, used widely in aeronautics and fluid dynamics \cite{Schlierenbook}. The method is also extensively used for the investigation of plasma processes in atmospheric gases \cite{Traldi2018} and, in particular for a wide range of processes involving laser induced plasma
\cite{Clayton1998, Iwase1998, Honda2000, Veloso2006, Batani2019}. Quite recently, ML techniques have been proposed to extract information from Schlieren imaging measurements of flows and shocks \cite{Znamenskaya2021, Cai2021, Ubald2022}. 

Plasma wakefield acceleration is a technology that promises a new generation of compact particle accelerators for scientific and commercial uses \cite{Joshi2003,Leemans2009}. Numerous research groups and collaborations are working worldwide to overcome the technological difficulties that wakefield acceleration poses. The AWAKE Collaboration hosted at CERN is a project where a high-energy proton bunch is used to drive plasma wakefields for electron acceleration \cite{Gschwendtner2016,Adli2018}. 
At the heart of the novel accelerator device, a 10-meter-long plasma channel achieves the modulation of the energetic proton bunch and the acceleration of witness electron bunches in the emerging wakefields. Created via photoionization using a terawatt laser system in a rubidium vapor source chamber \cite{Oz2014,Plyushchev2017}, plasma channel generation is in itself a complex problem of laser beam propagation/filamentation \cite{Couairon2007,Demeter2019,Demeter2021}. Optical diagnostic tools monitoring the plasma channel can thus have a significant role in optimizing, improving the accelerator device and understanding wakefield physics.   

In this paper, we consider using a Schlieren imaging setup as a diagnostic tool to determine vital parameters of a narrow plasma channel in tenuous ($\mathcal{N} = 10^{14}-10^{15} \mathrm{~cm}^{-3}$) atomic vapor. The setup is similar to the one tested to observe  atomic excitation in rubidium vapor \cite{Bachmann2018} and is geared toward determining the precise location and diameter of the rubidium plasma channel as well as the characteristic length for the spatial decay of plasma density. We test the use of Deep Neural Networks (DNNs) as universal function approximators to extract quantitative information on the plasma from the Schlieren images. We build datasets of simulated measurements to train networks with different architectures to estimate the parameters of the plasma.  We demonstrate that Schlieren imaging and machine learning techniques can be used effectively together to obtain information crucial for the operation of a proton-driven wakefield accelerator. 

\section{Schlieren imaging of a plasma channel cross-section}

\subsection{Measurement principle}
In the novel accelerator device of the AWAKE Collaboration, a 10-meter-long rubidium vapor source is used, with a TW power laser pulse propagating along the axis to ionize the vapor. Plasma for accelerator operation must satisfy very stringent constraints with respect to homogeneity of density. This is can be fulfilled by carefully tailoring rubidium vapor density in the chamber and achieving single-electron ionization of the atoms with a probability very close to unity \cite{Adli2019}. The propagation of the ultra-short, TW ionizing pulse along the vapor source is itself a complex nonlinear process \cite{Couairon2007}, especially because it is resonant with the rubidium $D_2$ transition line \cite{Demeter2019,Demeter2021}. Validating the quality of the plasma can be done near the downstream end of the vapor source, where a pair of observation ports on opposite sides of the chamber allow the passage of a probe beam transverse to the plasma channel axis. Precise measurements are hampered by the fact that plasma density distribution should be observed on a timescale much shorter than the $\sim 10\mu$s recombination/diffusion timescales and that the vapor (and hence the plasma) is extremely tenuous, its 
$\mathcal{N} = 10^{14}-10^{15} \mathrm{~cm}^{-3}$ density being 4-5 orders of magnitude smaller than the normal atmospheric density. 

\begin{figure}[htb]
\includegraphics[width=0.8\textwidth]{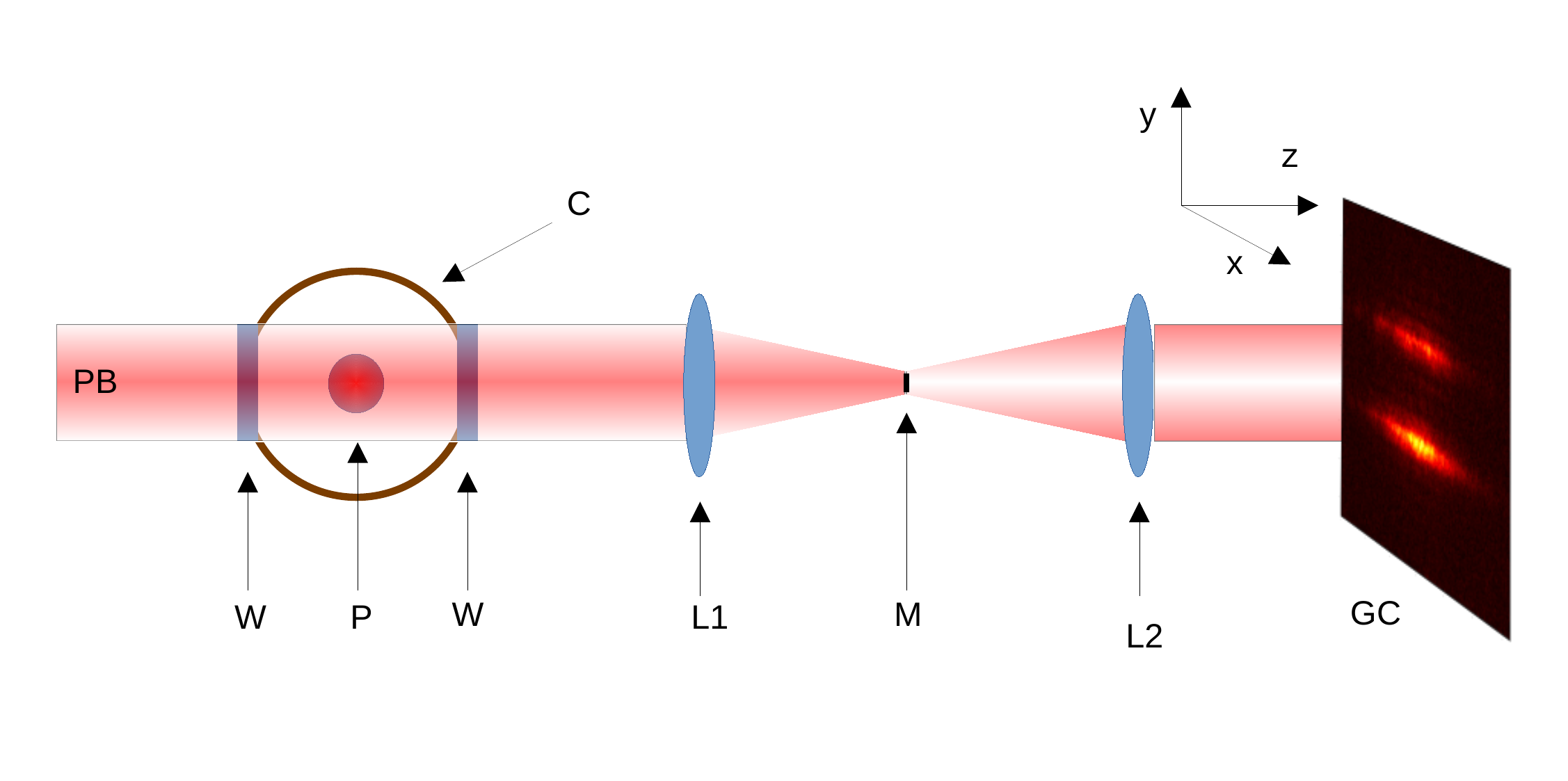}
\caption{Sketch of the Schlieren imaging measurement setup (not drawn to scale). PB - probe beam, C - vapor source chamber cross-section, W - viewport, P - plasma channel cross section, L1, L2 - 75 cm focal length lenses in 4f setup, M - mask, GC - gated camera.  }
\label{fig_setup}       
\end{figure}

The measurement setup that can be used for the required observation is sketched on Fig. \ref{fig_setup}. A Gaussian probe beam with beam waist parameter $w_0=2.6\mathrm{~mm}$ transits the chamber of the vapor source along the $z$ axis through a pair of viewports. The diameter of the chamber cross-section is 4 cm and the probe beam waist is positioned near the center of the chamber. Two lenses with focal lengths $f=75$ cm are placed in a '$4f$' setup \cite{SalehTeich} after the chamber, with a $D=1.5$ mm diameter circular mask positioned on the optical axis in the back focal plane of the first lens. A gated, image intensified camera (GC) detects the probe beam, triggered about 100 ns after the ionizing laser and gated to collect light for 100 ns exposure time. The probe beam is from a diode laser tuned to $\lambda=780.311$ nm, close to the $\lambda=780.241$ nm $D_2$ resonance wavelength of rubidium. 
With this choice, the anomalous dispersion around the resonance line yields a refractive index change of 
$\delta n_v=10^{-4}-10^{-3}$ caused by the vapor for $\mathcal{N} = 10^{14}-10^{15} \mathrm{~cm}^{-3}$ densities. At the same time, refractive index change due to the plasma dispersion is $\delta n_p=\sqrt{1-\omega_p^2/\omega^2}-1=(-2.7)\cdot10^{-8}$ \textemdash $(-2.7)\cdot10^{-7}$, over 3 orders of magnitude smaller. 
The Rayleigh-length of the probe beam with these parameters is $R\approx 27$ m, so the transit of the probe beam across the 4 cm thick layer of rubidium vapor amounts to a phase-shift of the Gaussian beam profile due to vapor dispersion, combined with an attenuation factor due to absorption. The circular mask at the focal plane between the two lenses acts as a high pass filter that blocks all of the probe light, unless a plasma channel (the Schlieren object) modulates the probe beam phasefront sufficiently such that some of the probe is deflected around the mask's edge. 

\begin{figure}[htb]
\includegraphics[width=0.8\textwidth]{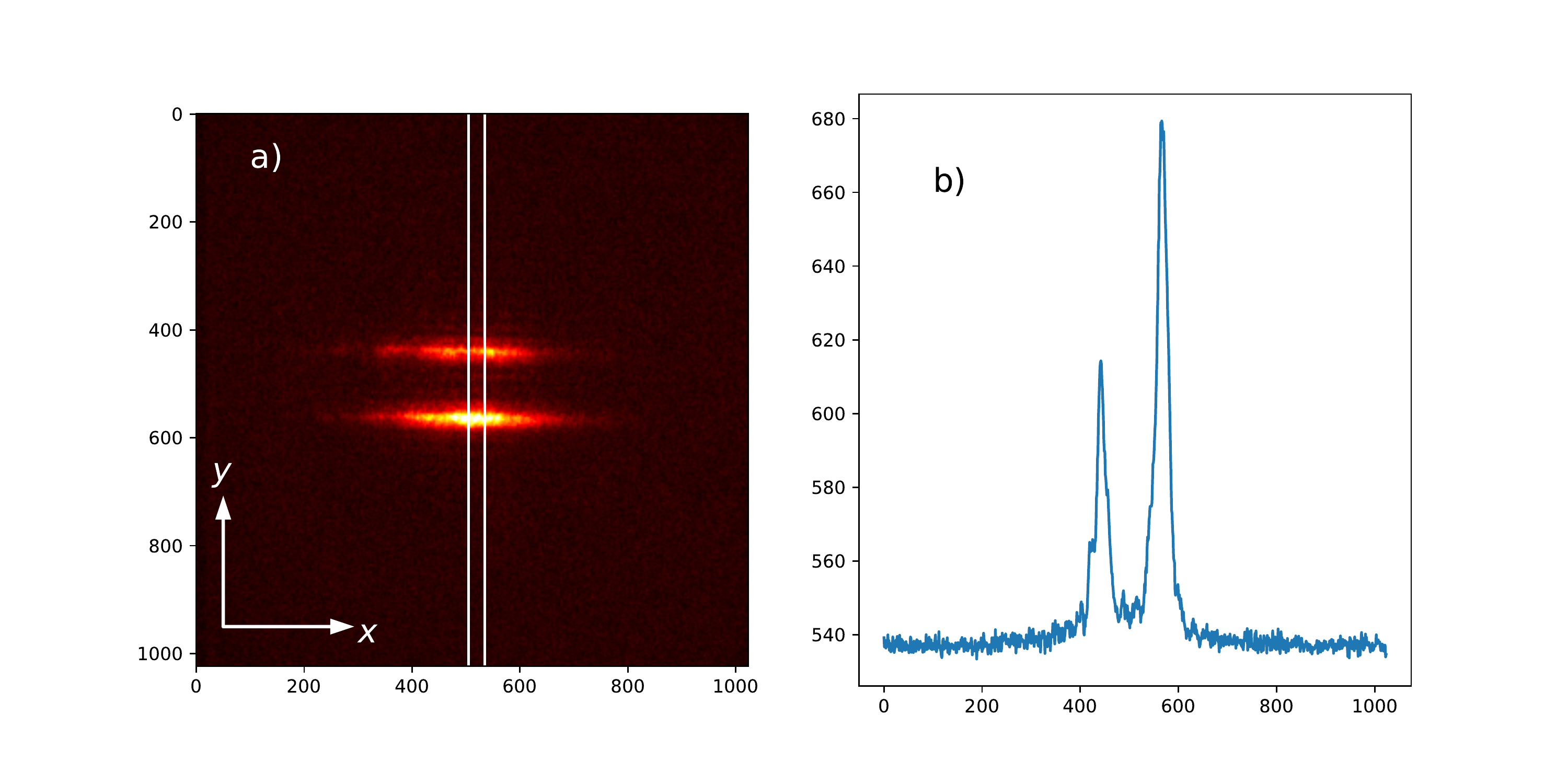}
\caption{a) Schlieren image on the gated camera. Lines in the middle mark the region of interest (ROI) from which we calculate the lineout around the probe beam center. b) Lineout taken from the Schlieren image ROI by averaging along $x$.}
\label{fig_schlieren}       
\end{figure}

An example of a measured image can be seen on Fig. \ref{fig_schlieren} a). Now the probe beam diameter ($\sim$mm) is negligible with respect to the spatial scale ($\sim$m) at which the plasma channel cross-section changes along the $x$ axis (the direction of propagation for the TW ionizing pulses). Therefore, the refractive index variation (and hence probe beam phase modulation) changes only along the $y$ direction. Changes along the $x$ direction on the measured image are only due to the variation of the probe beam amplitude along this coordinate. Thus the measured image contains stripes parallel with the $x$ axis and it is convenient to create a 1D lineout along $y$ by taking a region of interest (ROI)  from the region of $x$ where the probe beam is the most intense and averaging this along $x$ (Fig. \ref{fig_schlieren} b) ). This procedure helps reduce the noise level of the signal and produces a 1D curve that contains the same information on the plasma as the 2D camera image does.  The task is to extract quantitative information on the plasma channel from this curve.   

\subsection{Obtaining plasma parameters}
\label{subsec:plasmaparams}

To help interpret images obtained in the measurement, we first note, that if the plasma distribution in the $y-z$ plane is given, we can easily calculate the measured signal. The precise refractive index and absorption parameter for the probe beam wavelength can be obtained from the composite lineshape function using the material parameters of the rubidium $D_2$ line \cite{Siddons2008}. (Note that the vapor densities used here require that we augment the description of \cite{Siddons2008} with a pressure broadening term in the homogeneous lineshape \cite{vanLange2020}.) The integrated phaseshift of the probe beam and the overall attenuation can then be computed, and the transit across the $4f$ system with the mask can be calculated using the standard formulas for Fourier optics \cite{SalehTeich}. 
Therefore we start the analysis by assuming some functional form for the ionization probability in the vapor,
and observing the {\em simulated} signal that this plasma distribution would produce. 

To find a physically meaningful set of functions for the plasma density, we note that for relatively small ionizing pulse energies, we expect the ionization probability to be some power of the pulse intensity in general for multiphoton ionization. For large ionizing pulse energies, ionization probability saturates to values very close to unity in the central part of the beam \cite{Adli2019,Demeter2019}. The plasma channel is assumed to be axisymetric in the $y-z$ plane, with center relatively close to the axis of the vapor source (and optical axis). Since a shift of the plasma in the $z$ direction (parallel to the probe beam propagation) cannot be detected by the setup (the quantities we measure arise as integrals along $z$), we characterize the plasma center location with a single coordinate $y_0$, measured from the optical axis. The plasma density is thus assumed to have the following form:
\begin{equation}
\label{eq:plasma-density}
 \mathcal{N}_{plasma} = \left\{ 
 \begin{aligned}
 &\mathcal{N}_0 P_{max}, \mathrm{~if~} r\leq r_0,\\
 &\mathcal{N}_0 P_{max}\exp\left(-\frac{(r-r_0)^2}{t_0^2}\right) , \mathrm{~if~} r>r_0.
  \end{aligned}\right.
\end{equation}
Here $\mathcal{N}_0$ is the vapor density, $P_{max}$ is the maximum ionization probability of the vapor at the plasma channel center, $P_{max}\in [0,1]$.
$r=\sqrt{(y-y_0)^2+z^2}$ is the geometric distance from the plasma channel center, located at $(y,z)=(y_0,0)$, $r_0$ is the radius of the plasma channel core where the ionization fraction (and hence the plasma density) is constant, and finally $t_0$ is the parameter that characterizes the width of the transition region between the plasma channel center and the completely unionized vapor. 

\begin{figure}[htb]
\includegraphics[width=\textwidth]{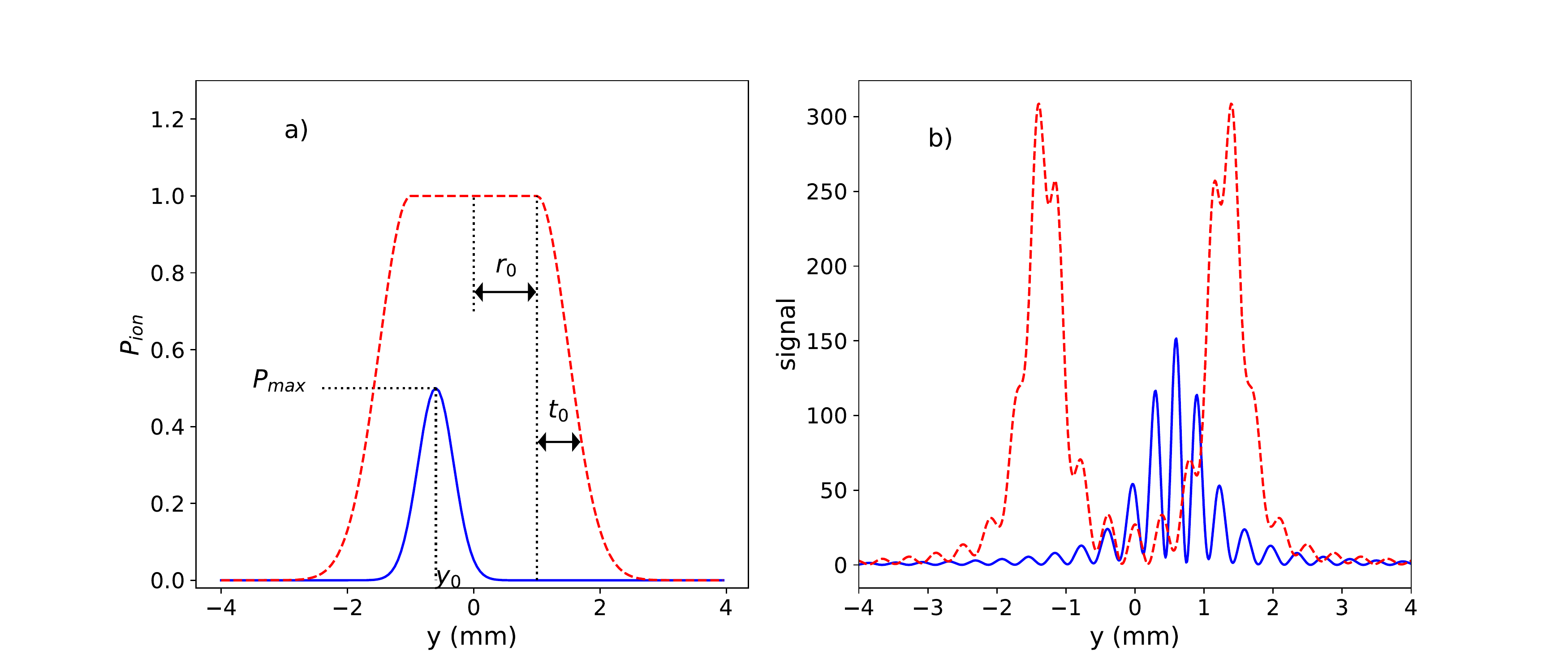}
\caption{a) Ionization probability for a weakly ionized ($P_{max}<1, r_0=0$), narrow plasma channel (solid blue line) and a saturated core $P_{max}=1, r_0>1$, wide plasma channel (dashed red line). b) The corresponding simulated signals (signal for a weakly ionized plasma has been scaled up for better visibility). }
\label{profiles}       
\end{figure}

The functional form written here can account for a weakly ionized vapor, when $r_0=0$ and the plasma density distribution is an axisymmetric Gaussian. It can also account for the opposite case, when there is a sizeable domain of fully ionized vapor $P_{max}=1, r_0>0$ and a Gaussian shaped transition region surrounding it (see Fig. \ref{profiles} a)). Clearly, not all parameter sets are physically realistic. Since the central, constant density region is associated with a saturation of the ionization fraction, $r_0>0$ happens only for $P_{max} \approx 1$.        

Given the above explicit functional form, we can calculate simulated signals on the gated camera for any set of 
plasma parameters $\{y_0,r_0,t_0,P_{max}\}$, some examples can be seen on Fig. \ref{profiles} b). The mathematical task is now to determine the plasma parameters that had been used to give rise to the given (possibly noisy) signal. Because of the nonlinear, integral-type relationship between the parameters and the Schlieren signal, this is a difficult task. Therefore, in the following section we propose a novel method for processing the Schlieren signals with the application of DNNs. 

\section{Inferring the plasma parameters with neural networks}
\label{sec:dnn}

Machine learning techniques have been successfully utilized in many fields, where it is an essential necessity to provide a precise and quick evaluation of to the input data with significant non-linearities~\cite{russel2010}. A typical data-based application of a feedforward artificial neural network is the non-linear regression, which is aimed to infer some parameters from the input data:
\begin{equation}
    y_{j,Pred}=f(x)=A\left(\sum\limits_{i=1}^Nw_{ij}x_i +b_i\right),
\end{equation}
where $A$ is some non-linear activation function, $N$ is the number of the neurons in the layer, $b_j$ is a bias vector and the $w_{ij}$ matrix contains the trainable parameters. A network may consists of multiple such layers---this is the case when it is called a \textit{deep} neural network. During a supervised training cycle (\textit{epoch}) of the network, the training input data is evaluated and compared to a reference output according to well defined loss function, $\mathcal{L}(y_{Pred}, y_{True})$. The objective then is to minimize this loss function by optimizing the weights in the $w_{ij}$ matrices, which is performed by the backpropagation: the weights receive updates that are proportional to the partial derivatives of the loss function with respect to the weights. This process is then repeated iteratively in several epochs until some stopping condition.

During a supervised training of a DNN model, there are a variety of tunable parameters that are specific to the given architecture and training method (so called \textit{hyperparameters}), like the learning rate (which controls the extent of update that the weights receive during backpropagation), the moments of the stochastic gradient descent optimizer (like in the popular Adam algorithm~\cite{kingma2017adam}) or some weight parameters in a multi-component loss function~\cite{Dosovitskiy2020You}. However, one of the most crucial and necessary element of the training is undoubtedly a good quality training dataset.

\subsection{Dataset generation}
\label{subsec:dataset}

Based on the formulation discussed in \ref{subsec:plasmaparams}, we have implemented a Python script that is able to simulate the Schlieren signal from a given set of plasma parameters. We have utilized this script to generate datasets for the training, validation and testing of the DNNs.
The datasets contained the simulated Schlieren signals sampled at 1024 points, paired with the set of generating plasma parameters $\{y_0,r_0,t_0,P_{max}\}$ used to obtain them. For each set of parameters, $y_0$ and $t_0$ were randomly chosen, distributed uniformly on a given interval. $P_{max}$ was chosen such that its $1/n$-th power was a uniform random number on the interval $[P_0, 1]$:
\begin{equation}
	P_{max} \in [ P_{0}^{n}, 1 ]^{1/n}
\end{equation}
This method skews the probability distribution of $P_{max}$ to favor values close to 1 somewhat, the exact amount depending on the positive integer generating parameter $n$. Note that $n=1$ corresponds to uniformly distributed $P_{max}$ values on the $[ P_{0}, 1 ]$ interval. For $r_{0}$, we enforced the following rule:
\begin{equation}
	r_{0} < \frac{Q}{(1 - P_{max})^{2}}
\end{equation}
with $Q = 0.25 \, \mathrm{\mu m}$. This value guarantees $r_{0}$ to have substantial values only when $P_{max}$ is close to 1.

In order to perform the training process with physically realistic data, the signal was slightly smeared with additive and multiplicative Gaussian noises. 
We then filtered the signals to reject samples whose signal to noise ratio was judged too small to evaluate reliably. First, we dropped samples whose maximum amplitude was less than $5.0$ units. The reason for this is that, such samples resemble only noise, and no peaks or interference patterns can be extracted from them. Then, we dropped samples with the absolute value of their mean less than $1.0$. Our argument behind it is that there are samples with a maximum amplitude greater than $5.0$ resembling noise and having a single large value somewhere, however, further raising the minimum accepted signal amplitude would exclude samples with acceptable signal/noise ratio. Note that the filtering also affects the statistical distribution of generating parameters in the final dataset created. Parameter sets that yield a plasma that is "undetectable" at the given noise level - e.g. because the ionization probability is too small, or the plasma is shifted too far out of the probe beam to be detected - are excluded.  This filtering of the data greatly improves the performance of the network, but it does not introduce any artificial, unwanted bias. 

\begin{figure}[htb]
\includegraphics[width=0.49\textwidth]{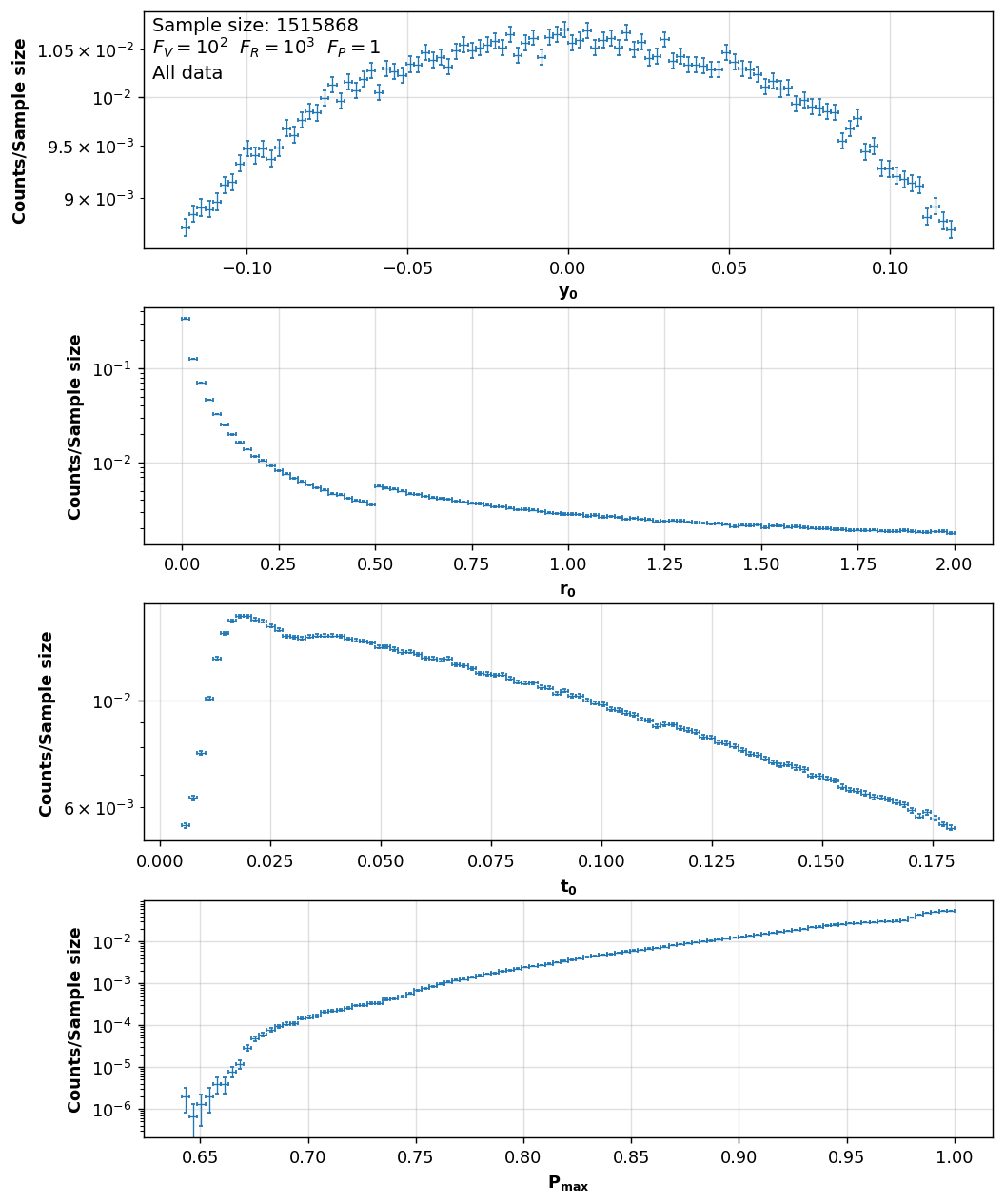}
\includegraphics[width=0.49\textwidth]{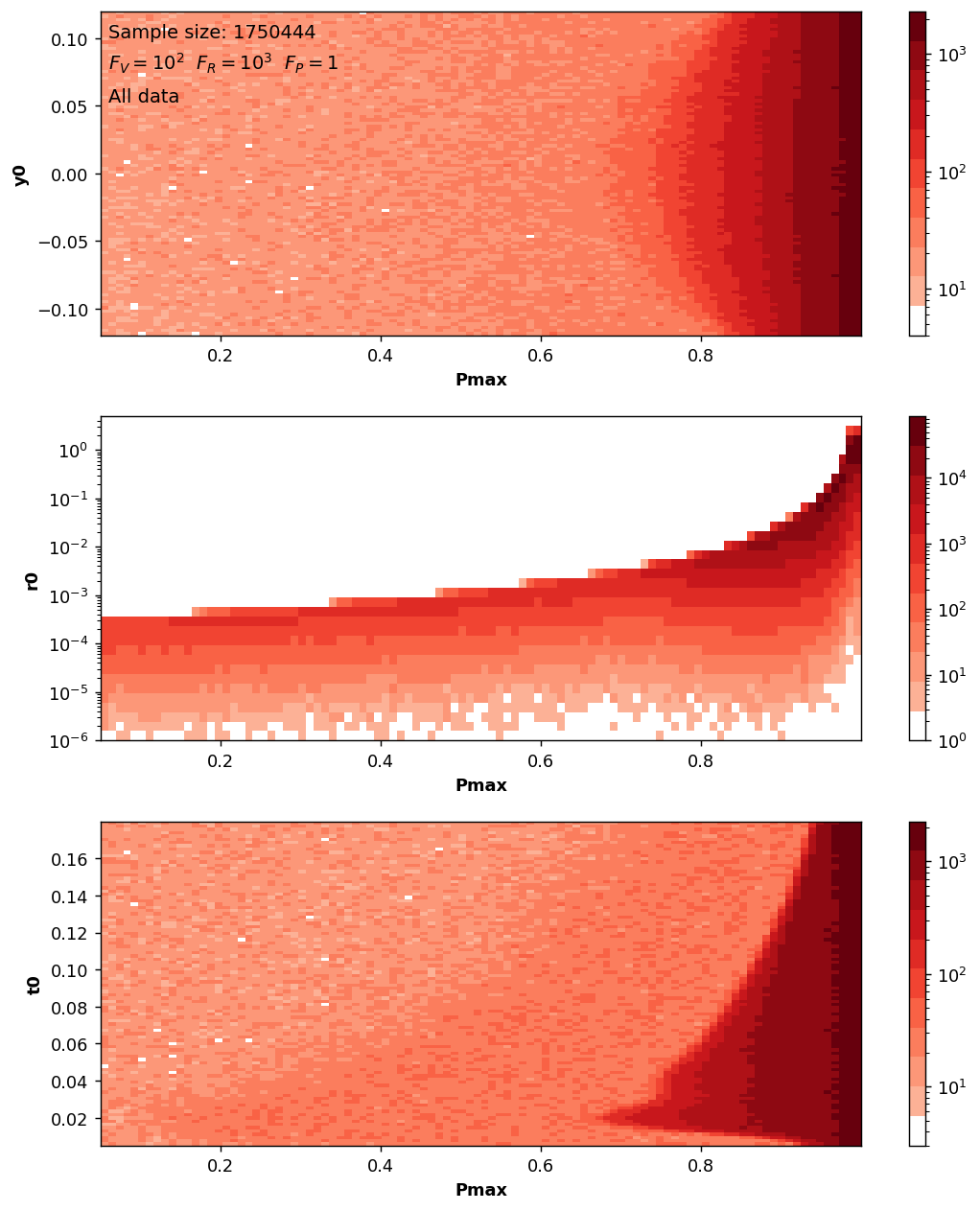}
\caption{The distribution of the parameters in the training data.}
\label{fig:sampleDistr}
\end{figure}

In most of the machine learning applications, it is crucial to standardize the data with some pre-processing method. However, in our case, we have full control over the generation of the simulated datasets, therefore only the followings have been considered:

\begin{enumerate}
    \item The input is a vector of 1024 elements, representing the detector image. Since the signal amplitude is sensitive to the degree of ionization, we did not apply any scaling to the input.
    \item In order to improve the learning process and reduce numerical instabilities, the plasma parameters have been multiplied with a constant factor to scale them into an approximately uniform range:
    \begin{subequations}
    \begin{align}
    	\tilde{P}_{max} & = F_{P} \cdot P_{max}, \\
    	\tilde{y}_{0} & = F_{Y} \cdot y_{0}, \\
    	\tilde{t}_{0} & = F_{T} \cdot t_{0}, \\
    	\tilde{r}_{0} & = F_{R} \cdot r_{0}.
    \end{align}
    \end{subequations}
\end{enumerate}
    The scaling factors have been chosen as: $F_{P} = 1$, $F_{Y} = F_{T} = 10^{2}$ and $F_{R} = 10^{3}$. The overall distributions of the parameters after these scalings are plotted on Figure \ref{fig:sampleDistr}.

Both for training and validating purposes, we generated realistic and noiseless datasets with different distributions given by $n = 1, 3, 5, 7$ and $10$, respectively. First, we trained and validated the networks on distinct datasets, that is, using datasets corresponding only to $n=1$ with noise, then e.g.~$n=5$ with noise, etc. We accepted a configuration only if the validation was convincing both for realistic and noiseless datasets. We also tested cross-validation in terms of noise and distribution in $P_{max}$. That is, we tried to validate a network trained with realistic data on noiseless data and vice versa, and also tried validating a network such that the training and validating datasets have different distributions in $P_{max}$. In theses cases, we experienced bad correlations between the true and the predicted parameters. According to theses experiences, first we decided to use unified datasets both for training and validating purposes, i.e.~the datasets contained realistic and noiseless samples from all $P_{max}$ distributions, given above. Later on, we used only realistic training and validating datasets with mixed distributions in $P_{max}$.

The unified and filtered datasets contained a total amount of $1.40M$ training and $0.23M$ validating samples, respectively.

\begin{figure}[htb]
\includegraphics[width=0.73\textwidth]{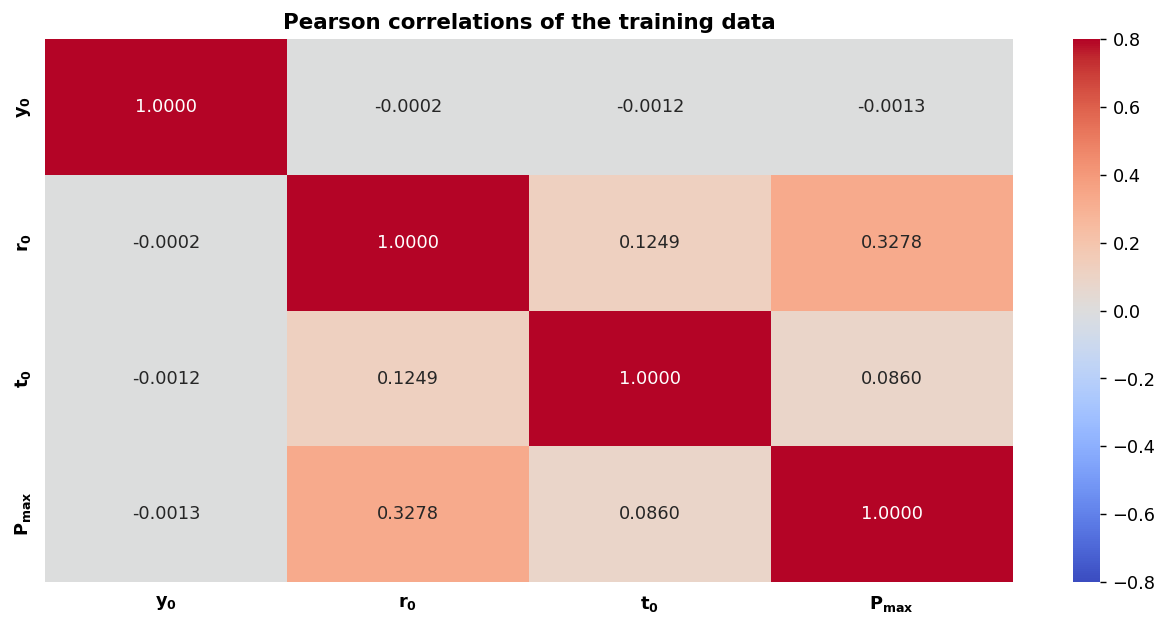}
\caption{The Pearson correlations in the training data.}
\label{fig:train_pearson}
\end{figure}

According to the argument in \ref{subsec:plasmaparams}, there is a strong correlation between the parameters $P_{max}$ and $r_0$: the $r_0>0$ values are favored only when $P_{max}\sim1$. In order to improve the statistics of the training dataset, an additional set was also generated, that allowed configurations only with $r_0>0.5$ mm---hence there is a small jump in the $r_0$ distribution of Figure \ref{fig:sampleDistr}. The correlations between the parameters are also visualized in Figure \ref{fig:train_pearson}, which presents the $r_{xy}$ Pearson correlation coefficients of the training data with a population size of $n=1,515,868$ samples, defined by the following equation:
\begin{equation}
  r_{xy}=\frac{\sum\limits_{i=1}^{n}(x_i-\bar{x})(y_i-\bar{y})}{ \sqrt{\sum\limits_{i=1}^{n}(x_i-\bar{x})^2} \sqrt{\sum\limits_{i=1}^{n}(y_i-\bar{y})^2} }
\end{equation}

As the figure shows, the $r_0$ (the radius of the plasma) is strongly correlated with the maximum ionization probability, while the transition region width, $t_0$ is only slightly correlated with $P_{max}$ and $r_0$. The $y_0$ location is basically uncorrelated with the other parameters.

\subsection{Methodology}

\begin{figure}[htb]
\begin{center}
  \includegraphics[width=0.20\textwidth]{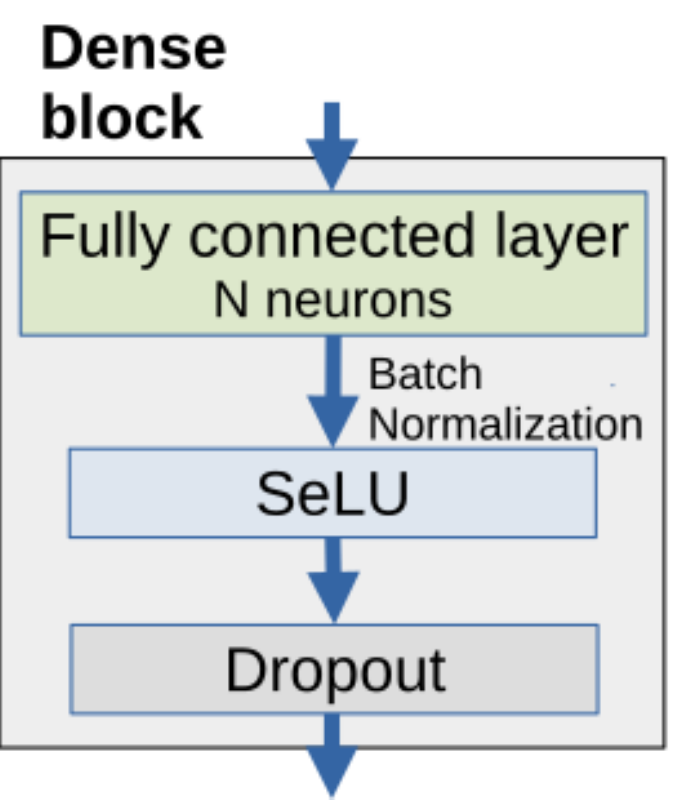}
  \caption{A basic building block of the applied neural networks.}
  \label{fig:block}       
\end{center}
\end{figure}

Our aim is to develop a robust framework that is able to infer the plasma parameters from the Schlieren signals with high accuracy. To achieve this, a customizable 
DNN framework has been implemented in Python, using Keras v2.7.0 with Tensorflow v.2.7.0 backend~\cite{chollet2015keras,abadi2016tensorflow}. The basic building block of the framework, referred as \textit{Dense block} is sketched on Figure \ref{fig:block}, which consists of a fully connected layer, followed by a batch normalization, a scaled exponential linear unit (SELU  \cite{https://doi.org/10.48550/arxiv.1706.02515}) and a dropout layer with fixed dropout rate of 0.1.


\begin{figure}[htb]
\includegraphics[width=0.93\textwidth]{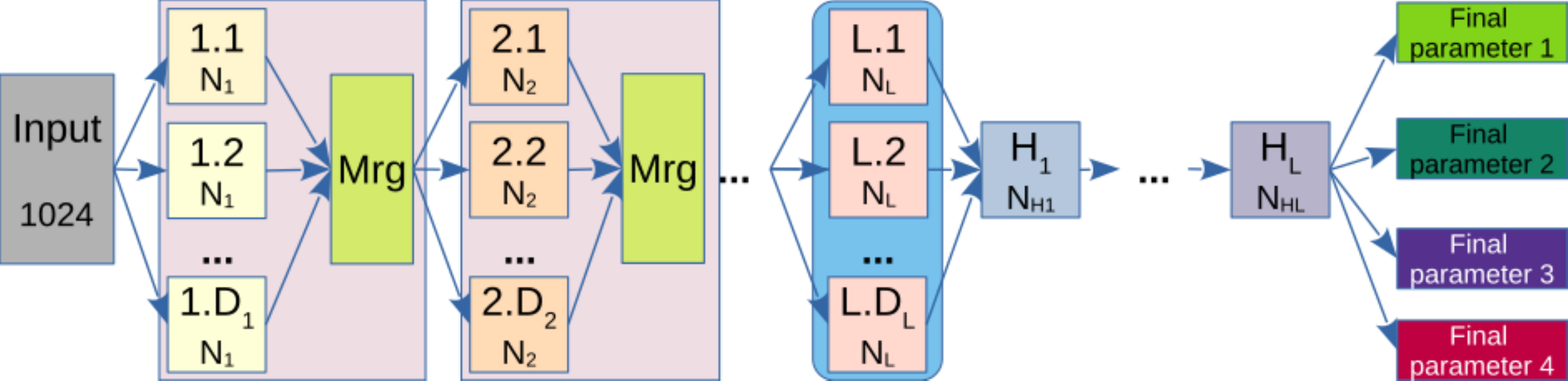}
\caption{The structure of the implemented neural networks.}
\label{fig:threads}
\end{figure}

Our neural networks have been composed of such building blocks as it is depicted on Figure \ref{fig:threads}, which has three distinct parts and several configurable parameters. In the first part (marked with light pink background) one or more feature extraction block process the input data, with $D_i$ parallel Dense blocks and with $N_i$ neurons in each block. Subsequently, the output of the $D_i$ Dense blocks are merged, which is then followed by a SoftMax activation. The $L$ index specifies the $L-1$ number of the consecutive feature extraction blocks, and afterwards a concatenation layer merges the parallel blocks in the $L^{th}$ layer (as an analogy for flattening, marked with light blue background). In the third part, $H_L$ hidden layer follows, with $N_{HL}$ neurons in the given layer. Finally, the last fully connected layer represents the four plasma parameters with linear activation. The mean absolute error has been chosen for the loss function, while the the optimization was performed with the Adam algorithm~\cite{kingma2017adam}. 
The initial learning rate has been slowly decreased with a linear decay.

The training, evaluating and testing were performed on the GPU clusters of the Wigner Scientific Computational Laboratory (WSCLAB). 

\subsection{Network implementations}
\label{subsec:networks}

In this study, several different architecture implementations have been investigated.  During training, the mean absolute error was utilized as the loss function, defined as:
\begin{equation}
    \mathcal{L}(y_{Pred}, y_{True})=\frac{1}{N}\sum\limits_{i=1}^N\left|y_{Pred}-y_{True}\right|
\end{equation}
Furthermore, the training process was monitored with the mean squared error and Log-Cosh errors as additional metrics~\cite{https://doi.org/10.48550/arxiv.2101.10427}, and  15\% of the training data was used for validating each epoch. Hereinafter, we refer to the calculation of the loss for every epoch as the measurement of training quality.
Obviously, bad convergence in the measurement of training quality shows bad correlation between the true and predicted parameters. However, it is still possible that with quick and clear convergence, the correlations are still poor. We found that it is sufficient to train the networks for $20,000$ epochs to get the best achievable quality of predictions.

Considering the initial learning rate, we tried different values in the $[10^{-4}, 10^{-2}]$ ranges. Taking lower initial values results in clearer and faster convergence in the measurement of training quality, however, with too small values, the correlations become bad. Taking too high initial values results in high fluctuations and very slow convergence in the measurement of training quality with bad correlations. We found that the best value for the initial learning rate is $3 \cdot 10^{-5}$.

We also tried different kinds of activation functions for the last layer in our networks, and found that the best choice is the linear activation, which is a common choice for regression tasks. Since in this way the possible values of the parameters were not limited neither from below nor from above, the network was forced to learn the physically relevant value ranges of the plasma parameters.

In Table \ref{tab:networks} the specific values of the configurable parameters and the final loss value for the listed variants are summarized. In the followings, the best three variants are compared.

\begin{table}[hbt]
  \centering
  \caption{The configurable hyperparameters.}
  \label{tab:networks}
  \begin{tabular}{p{0.26\textwidth}p{0.24\textwidth}p{0.24\textwidth}p{0.24\textwidth}}
    \hline
        &    \bf{Skirun} &  \bf{Bush} & \bf{LongBottle}\\
    \hline
    $D_L$ & 16, 8, 4 & 16, 4 & 1, 1, 1, 1, 1 \\
    $N_L$ & 512, 256, 128 & 256, 64 & 1024, 512, 256, 128, 64 \\
    $N_{HL}$ & 0 & 0 & 0 \\
    Trainable parameters & 9.6M & 4.3M  & 1.75M \\
    \hline
    Final loss &  0.00402 & 0.00667 & 0.00954\\
    \hline
  \end{tabular}
  \end{table}
  
Models named \textit{Bush}  and \textit{Skirun} included only feature extraction parts without additional hidden layers, while the model named \textit{LongBottle} is a special case of the network, where no explicit feature extraction blocks have been applied, therefore it is considered as a traditional deep neural network. For a reference of the models, see Table~\ref{tab:networks}. 

\section{Results}
\label{sec:results}

\begin{figure}[htb]
  \includegraphics[width=0.32\textwidth]{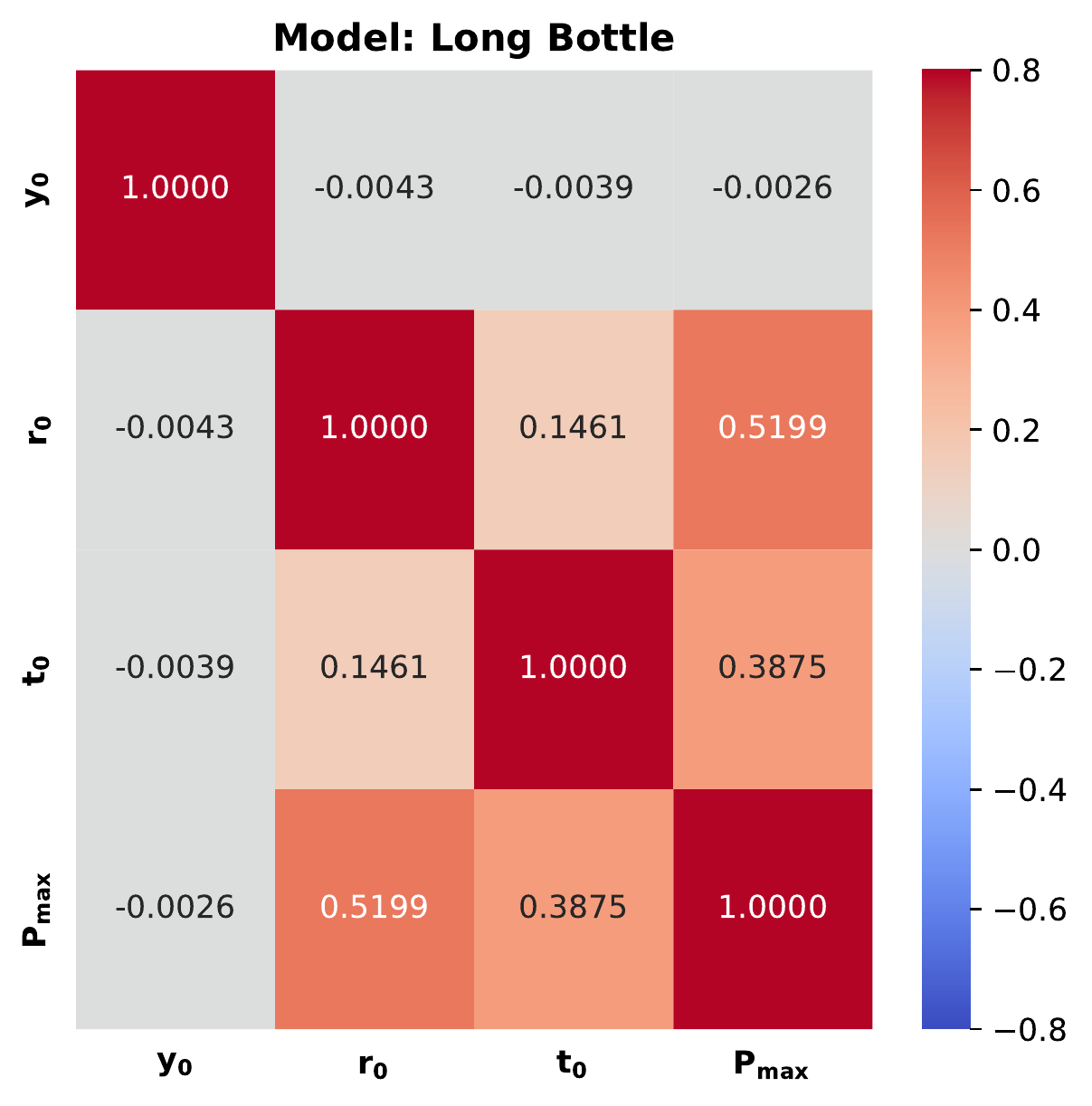}
  \includegraphics[width=0.32\textwidth]{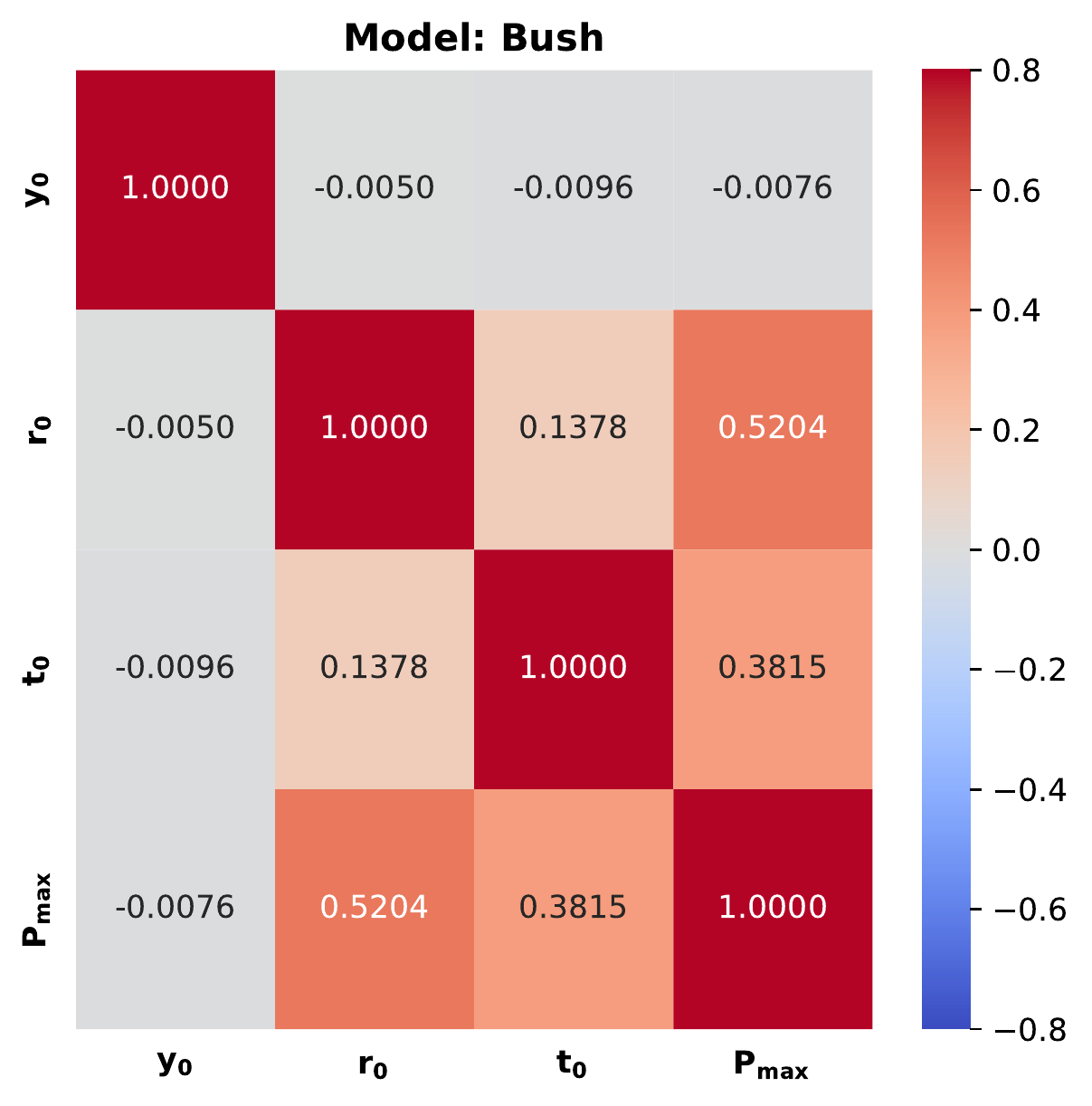}
  \includegraphics[width=0.32\textwidth]{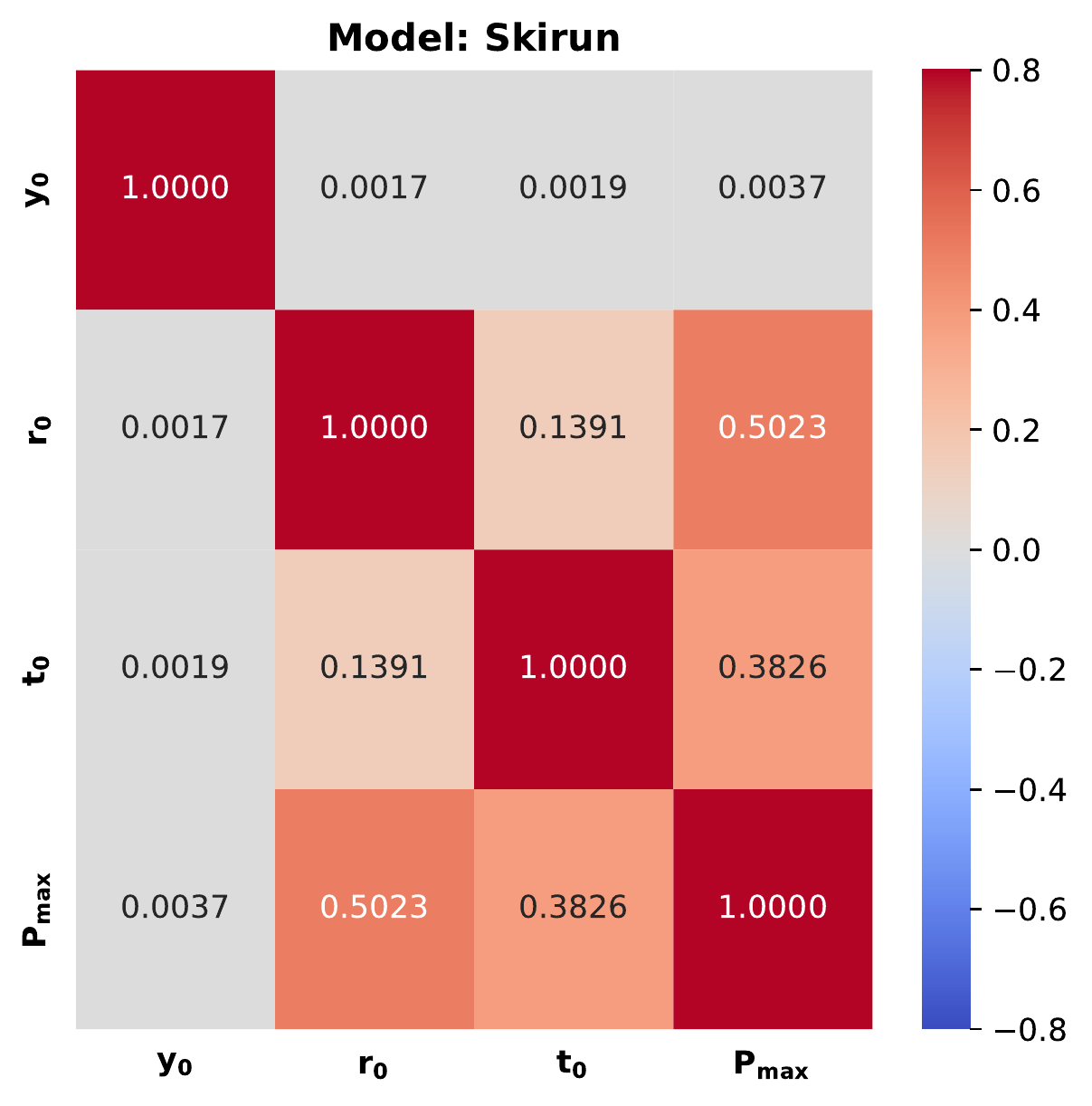}
  \caption{The learned Pearson correlations.}
  \label{fig:learned_pearson}
\end{figure}

In order to test the performance of the trained networks, the plasma parameters have been predicted from the validation dataset described in the previous section. Figure \ref{fig:learned_pearson} presents the Pearson correlation coefficients of the predicted parameters for the different networks.

The different architectures learned the same correlations, there is both qualitative and quantitative agreement. Comparing the results with the reference plotted in Figure \ref{fig:train_pearson}, the following conclusions can be drawn:

\begin{enumerate}
    \item the $y_0$ parameter is basically uncorrelated, which is well reproduced by all variants;
    \item the networks predict $\sim25\%$ higher $r_{t_0,r_0}$ correlation than the reference;
    \item the predicted $r_{P_{max},r_0}$ correlation is $\sim60\%$ higher, than the reference;
    \item in the reference, the $r_{P_{max},t_0}$ correlation in only $8.6\%$, while it is much higher, $\sim38\%$ from all the predictions.
\end{enumerate}

Note that the uncorrelatedness of the $y_{0}$ parameter means that the plasma parameters have a translation invariance with respect to $y_{0}$. This exactly agrees the fact that the properties of the plasma channel are independent of the exact position of the ionizing laser pulse if the  vapor has a uniform density distribution. It is also important to emphasize that we did not put any constraint on the plasma parameters, like $P_{max} \in [0,1]$. All the presented network variants recognized the physical range of the parameters along with the correlations between them, e.g.~larger values for $r_{0}$ are favored at higher values of $P_{max}$.

\begin{figure}[htb]
  \includegraphics[width=0.32\textwidth]{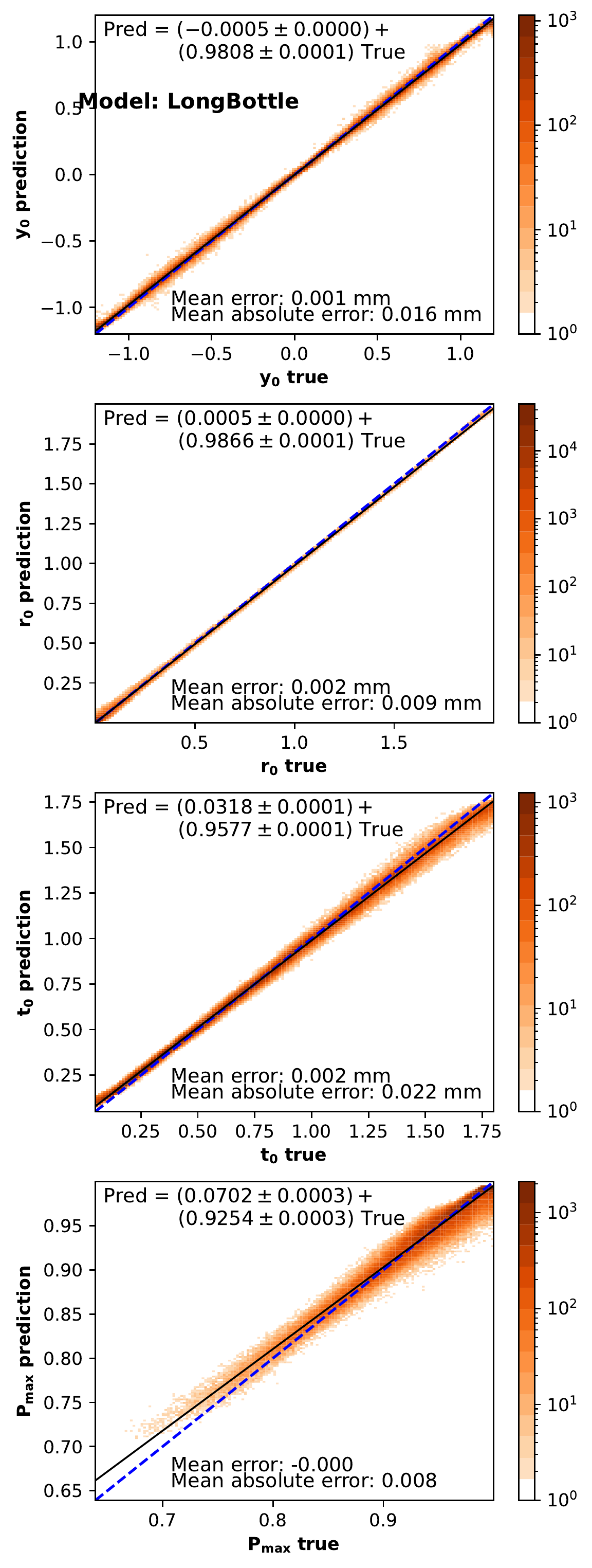}
  \includegraphics[width=0.32\textwidth]{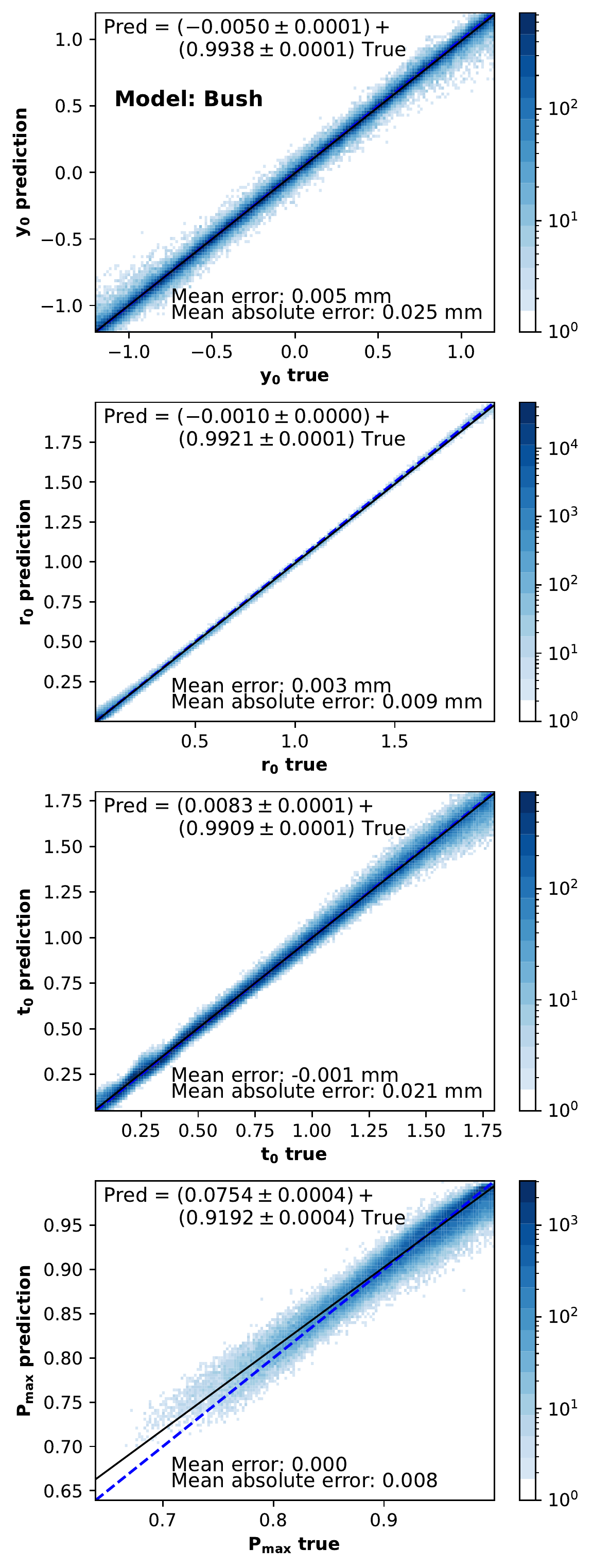}
  \includegraphics[width=0.32\textwidth]{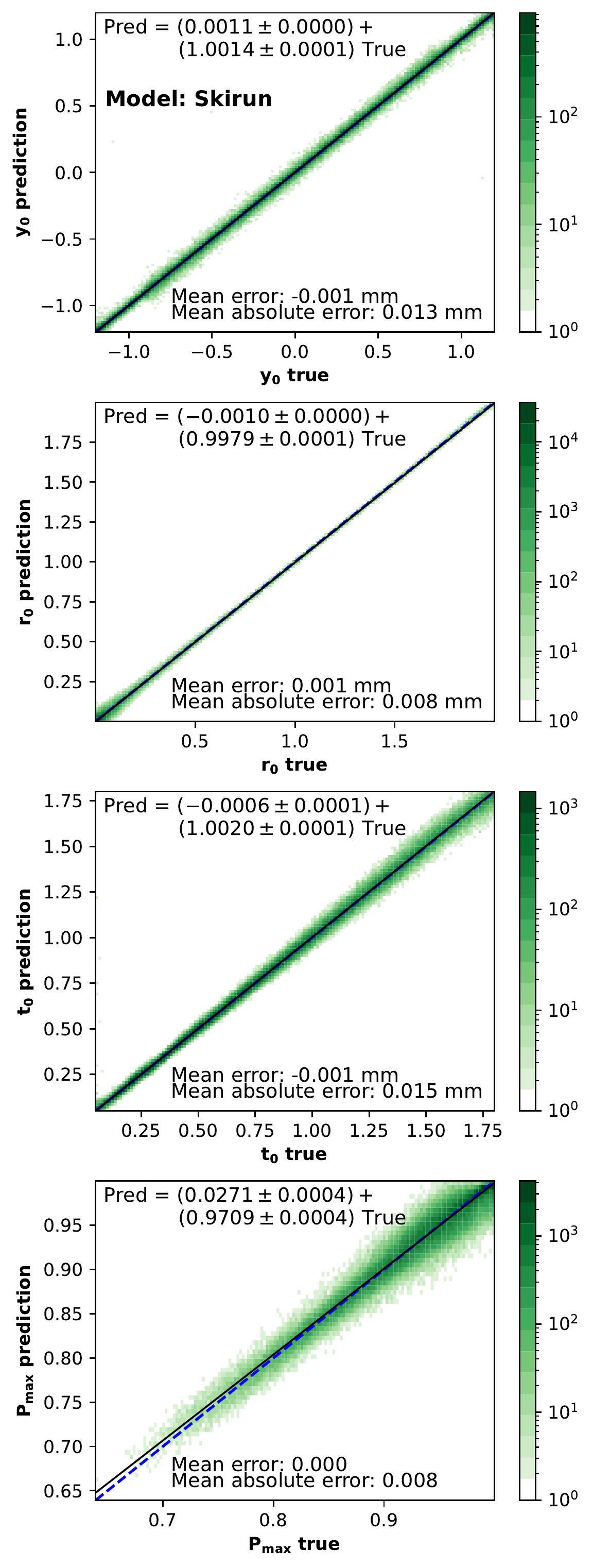}
  \caption{The learned parameter correlations.}
  \label{fig:learned_params}
\end{figure}

The predicted parameters are plotted versus the true values for all the investigated networks in Figure \ref{fig:learned_params}. All of the network variants produced a fairly adequate prediction of the plasma parameters---however, there are some significant differences. The \textit{LongBottle} variant produced the largest deviation for all parameters, but it is worthwhile to note that  among the investigated architectures this contained the least trainable parameters---only $1.75M$, which is less than the half of the second network. On the other hand, we have found that just increasing the number of trainable parameters is not enough for significant improvement, the feature extraction blocks are also necessary. 

To quantify the goodness of the predictions, first, a linear fit was performed to all parameters, and the mean error along with the mean absolute error is shown for all case. Second, the amplitude and phase errors were investigated according to the following considerations.

\begin{figure}[!htb]

\includegraphics[width=0.32\textwidth]{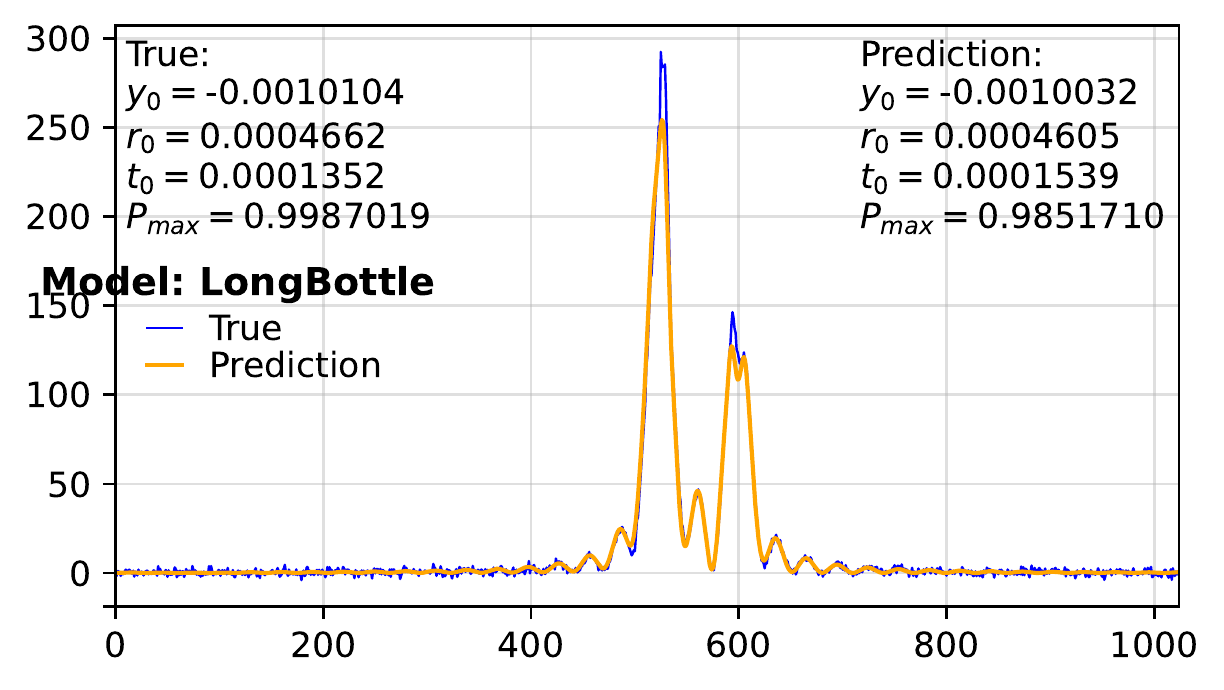}
\includegraphics[width=0.32\textwidth]{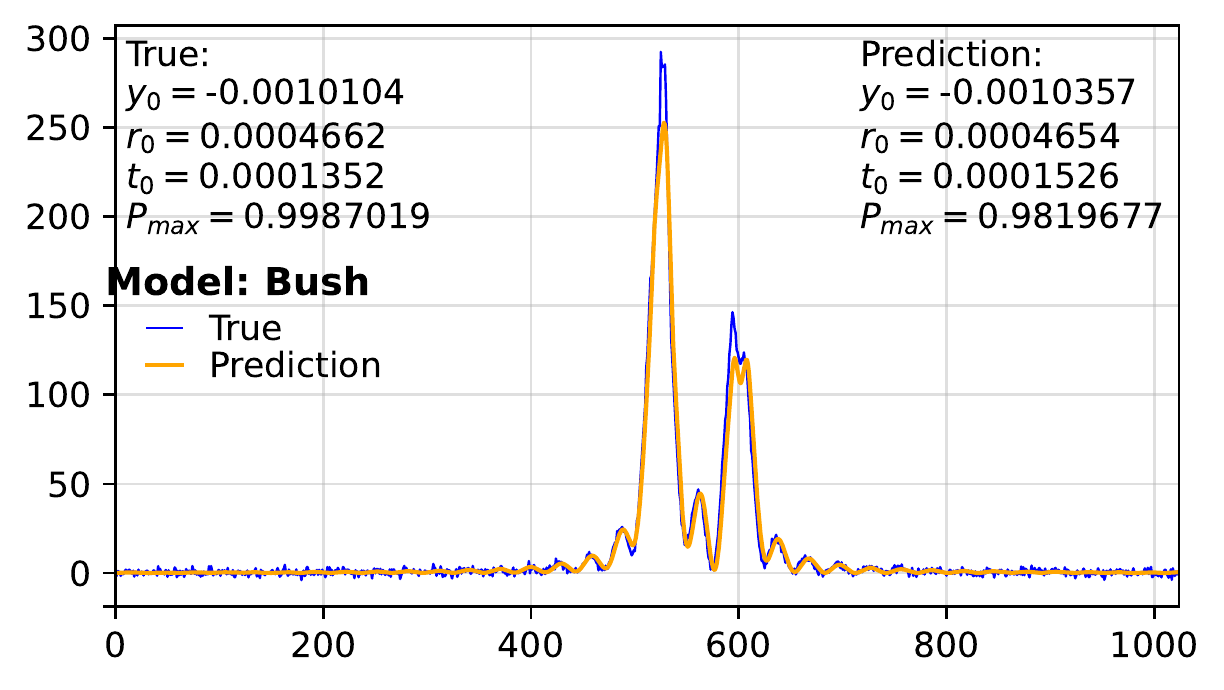}
\includegraphics[width=0.32\textwidth]{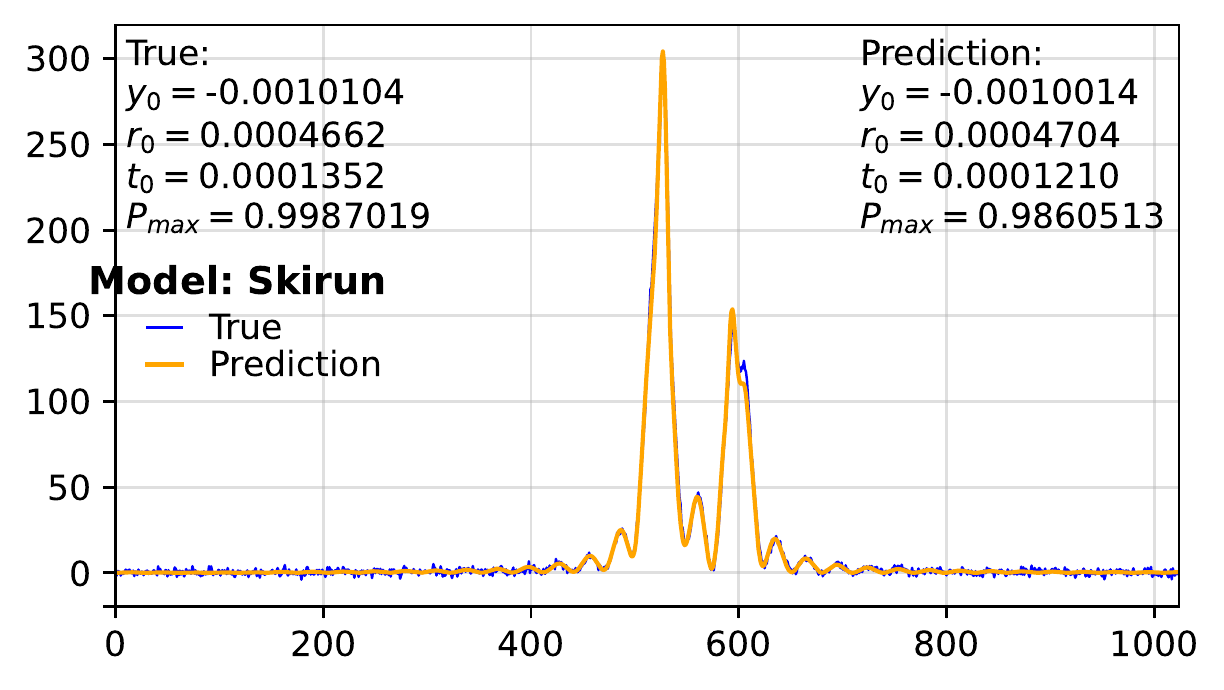}

\includegraphics[width=0.32\textwidth]{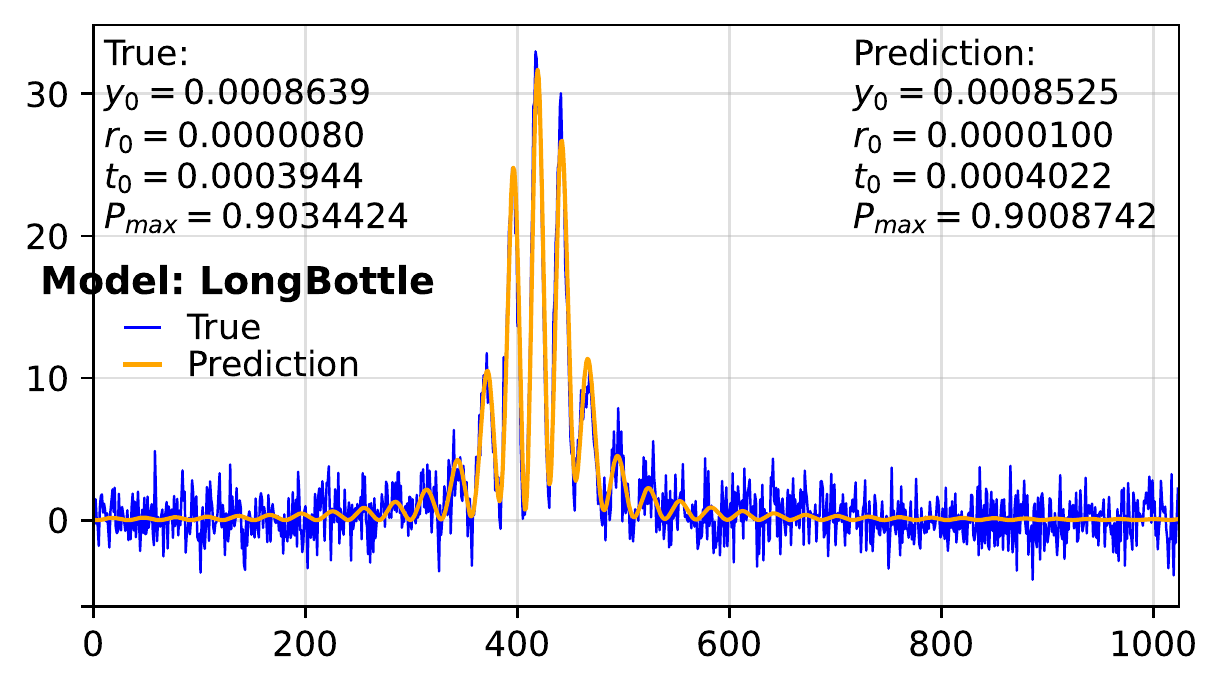}
\includegraphics[width=0.32\textwidth]{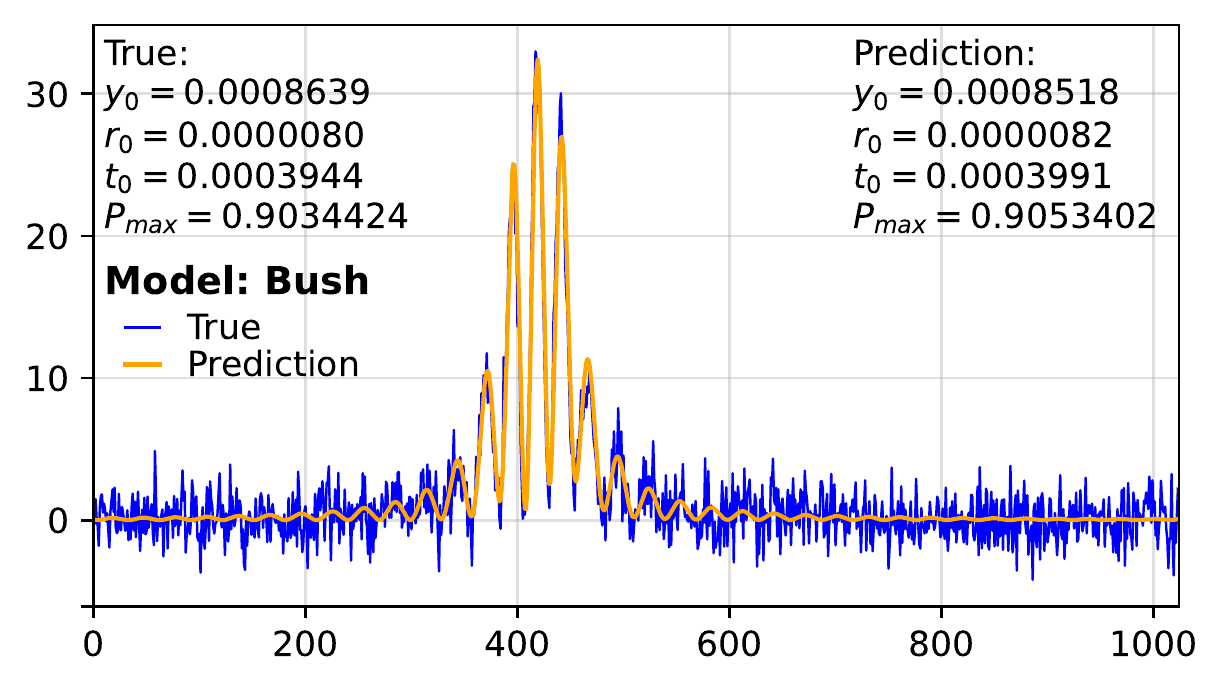}
\includegraphics[width=0.32\textwidth]{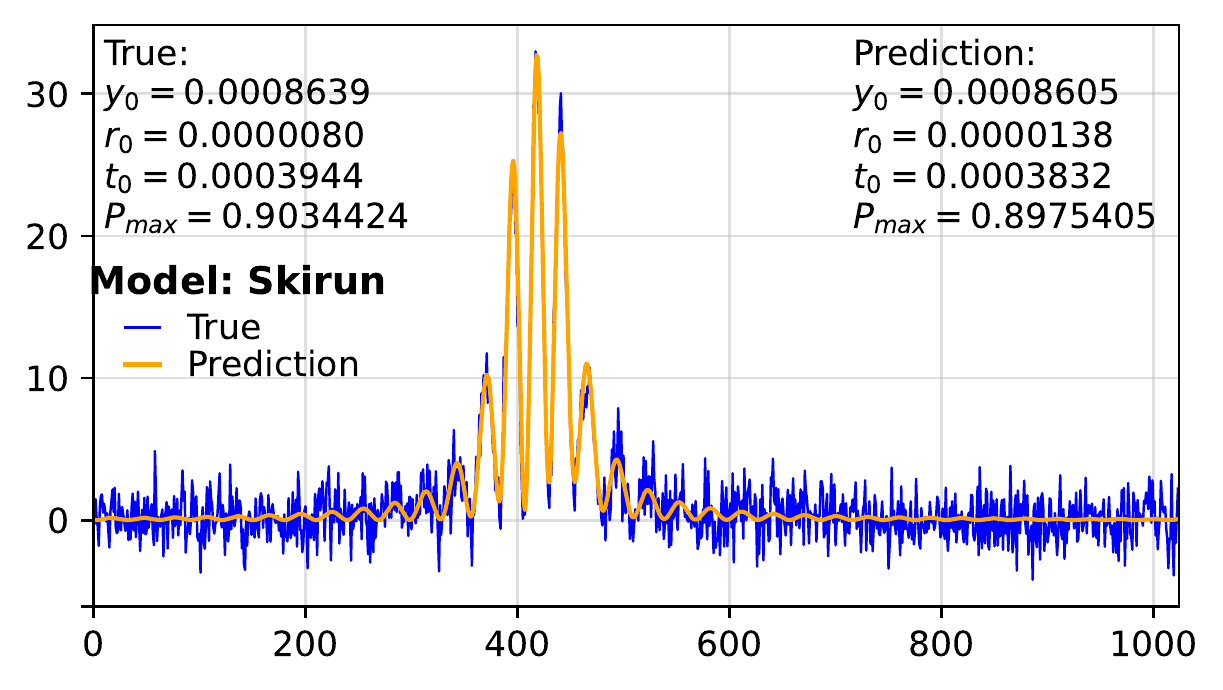}
\caption{Example predictions by the three models.}
\label{fig:examples}
\end{figure}

Let $y_{Pred}$, $y_{Noisy}$ and $y_{True}$ denote the predicted and the simulated Schlieren signals with and without the simulated noise, respectively. That is,
\begin{equation}
\label{eq:noisy}
	y_{Noisy} (x) \approx y_{True} (x) + n(x)
\end{equation}
with $n(x)$ denoting the noise such that
\begin{equation}
	\int_{x_{min}}^{x_{max}} n(x) \diff x = 0.
\end{equation}
Introducing the local noise to signal ratio
\begin{equation}
	\eta (x) = \frac{n(x)}{y_{True}(x)},
\end{equation}
equation \eqref{eq:noisy} can be rewritten as
\begin{equation}
	y_{Noisy} (x) \approx y_{True} (x) \bparenth{1 + \eta (x)}.
\end{equation}
Recall that for $P_{max} \approx 1$, the local noise to signal ratio can be neglected. However, the smaller values $P_{max}$ takes, the larger will be the fraction of the $\bparenth{x_{min}, x_{max}}$ interval where $\eta (x)$ plays a significant role. This can be seen on the top and bottom panels of Fig.~\ref{fig:examples}, respectively.

The high accuracy of our models suggests that after some slight phase and amplitude corrections, the true signals can be restored from the predicted ones. That is,
\begin{equation}
\label{eq:approx-accurate}
	y_{Pred} (x) \approx a \cdot y \parenth{x - x_{ph,corr}}
\end{equation}
with $x_{ph,corr}$ denoting the phase correction and $a$ being close to unity. $y (x)$ can either be $y_{Noisy} (x)$ or $y_{True} (x)$. As discussed above, for $P_{max} \approx 1$, the relation $y_{True} (x) \approx y_{Noisy} (x)$ holds. In order to see how the constant $a$ measures the quality of the fitting, first, consider the normalized signals in the followings:
\begin{equation}
	\tilde{y} (x) = \frac{y(x)}{C}
\end{equation}
with
\begin{equation}
	C = \int_{x_{min}}^{x_{max}} \abs{y(x)} \diff x.
\end{equation}
Let the phase dependent amplitude error defined as
\begin{equation}
	A_{err} \parenth{x_{ph}} = \int_{x_{min}}^{x_{max}} \abs{\tilde{y}(x) - \tilde{y}_{Pred} \parenth{x - x_{ph}}} \diff x
\end{equation}
with $\tilde{y} (x)$ being either $\tilde{y}_{True} (x)$ or $\tilde{y}_{Noisy} (x)$, and $x_{ph}$ being a small phase shift in the predicted signal. Due to the high accuracy of the predictions, we can assume that for some small value of $x_{ph,min}$, $A_{err} \parenth{x_{ph,min}}$ will be minimal. Therefore, we define $A_{err} = A_{err} \parenth{x_{ph,min}}$ as the amplitude error and $x_{ph,err} = x_{ph,min}$ as the phase error. Using the approximation showed in Eq.~\eqref{eq:approx-accurate}, it is easy to see that comparing the true, i.e.~noiseless signal to the predicted one, the expression of the amplitude error reduces to
\begin{equation}
	A_{err} = \abs{1 - a},
\end{equation}
meaning that $A_{err}$ is a good estimate for the relative error of the amplitude of the signal. When comparing the predicted signal to the noisy one, and considering the cases where the local noise to signal ratio is negligible, we get the same approximating expression for $A_{err}$ and therefore its meaning remains unchanged. However, when $\eta (x)$ is not negligible, $A_{err}$ neither does have such a nice graphical meaning as before, nor can be approximated with a simple formula. In contrast, when interpreting the phase error, we recall that both for the noiseless and noisy cases, the relation $x_{ph,min} = x_{ph,corr}$ holds. Therefore, $x_{ph,err} = x_{ph,corr}$, that is, the phase error is the slight correction with which the original signal can be restored from the predicted one with high accuracy.

We used the quantities defined above to measure the quality of the predictions as follows. For every test sample, we calculated $A_{err}$ and $x_{ph,err}$ and made a histogram from them. In the histogram of the amplitude error we divided the $\bparenth{0, A_{err,max}}$ interval to $1\,024$ bins, while in the histogram of the phase error we took $51$ bins since the phase error ranged from $-25$ pixels to $25$ pixels. We calculated the midpoints of the intervals representing the bins and assigned the counts of the bins to the corresponding midpoints. Then, we divided the number of counts of the bins with the total number of samples to obtain the probability of the signal having $A_{err}$ amplitude error and $x_{err}$ phase error, respectively. From these probabilities we calculated the mean values of $A_{err}$ an $\abs{x_{err}}$ with
\begin{equation}
	\avg{A_{err}} = \sum_{i} p \parenth{A_{err,i}} A_{err,i}
\end{equation}
and
\begin{equation}
	\avg{\abs{x_{err}}} = \sum_{i} p \parenth{x_{err,i}} \abs{x_{err,i}},
\end{equation}
respectively.

\begin{figure}[!htb]
  \includegraphics[width=0.32\textwidth]{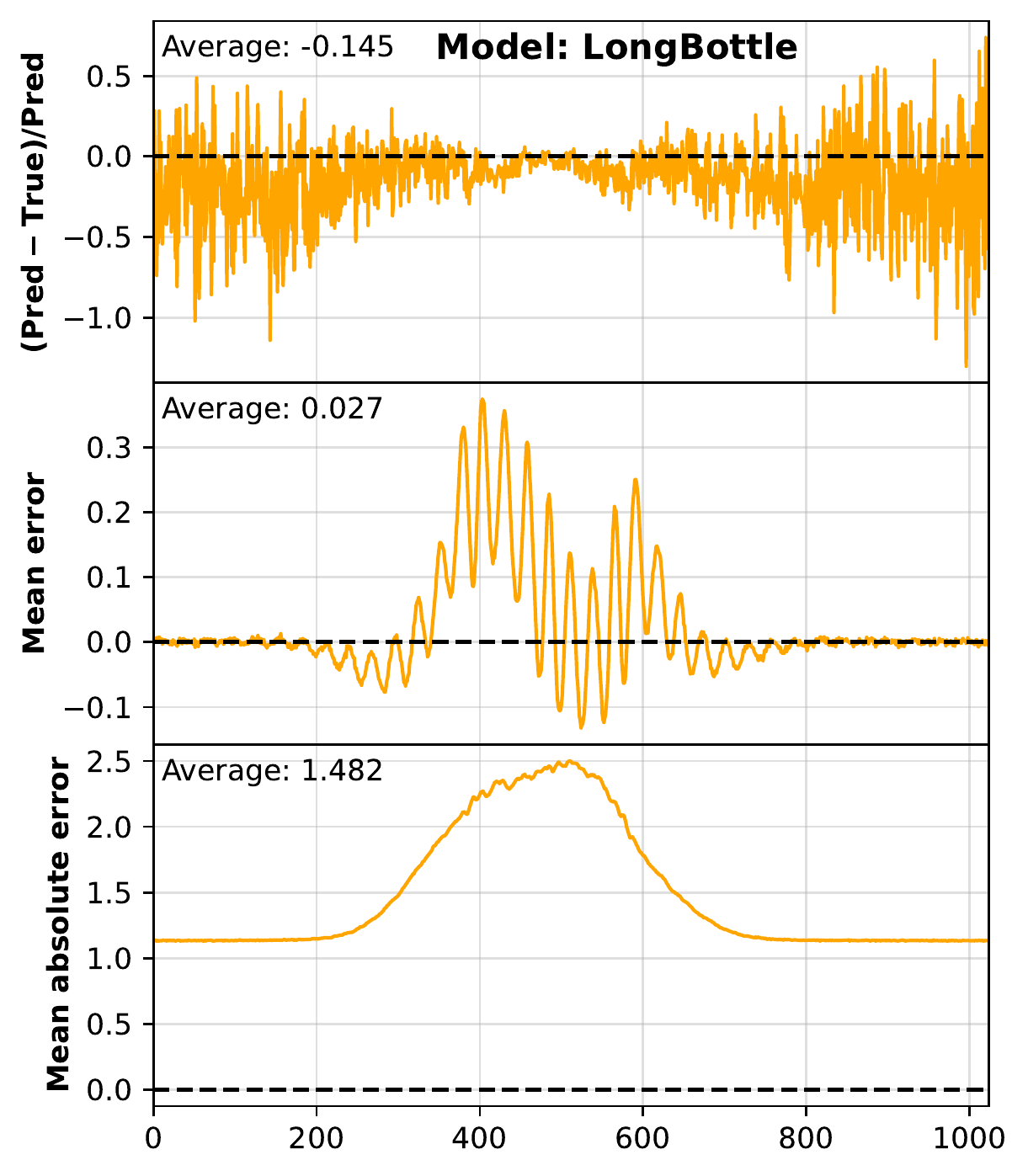}
  \includegraphics[width=0.32\textwidth]{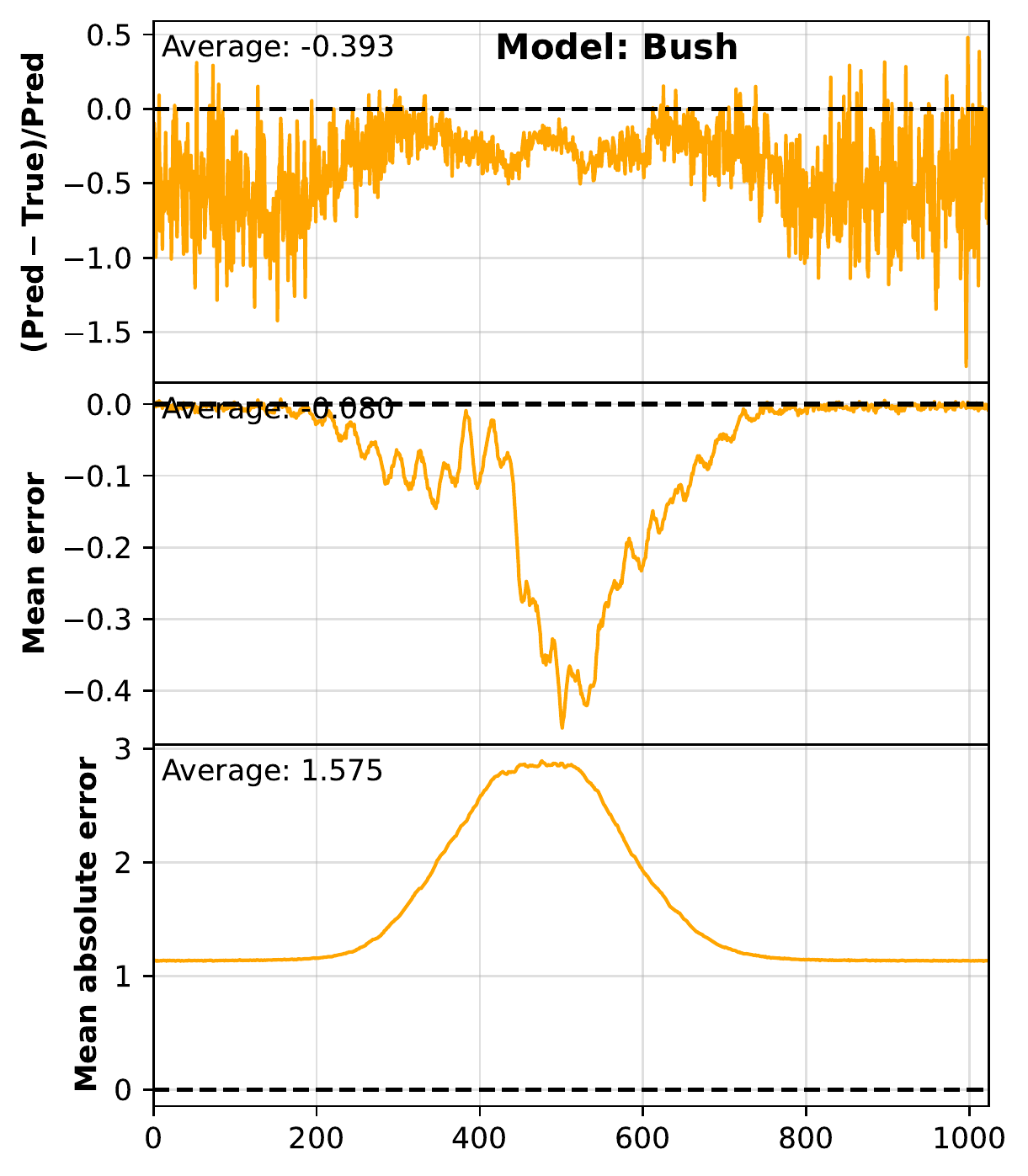}
  \includegraphics[width=0.32\textwidth]{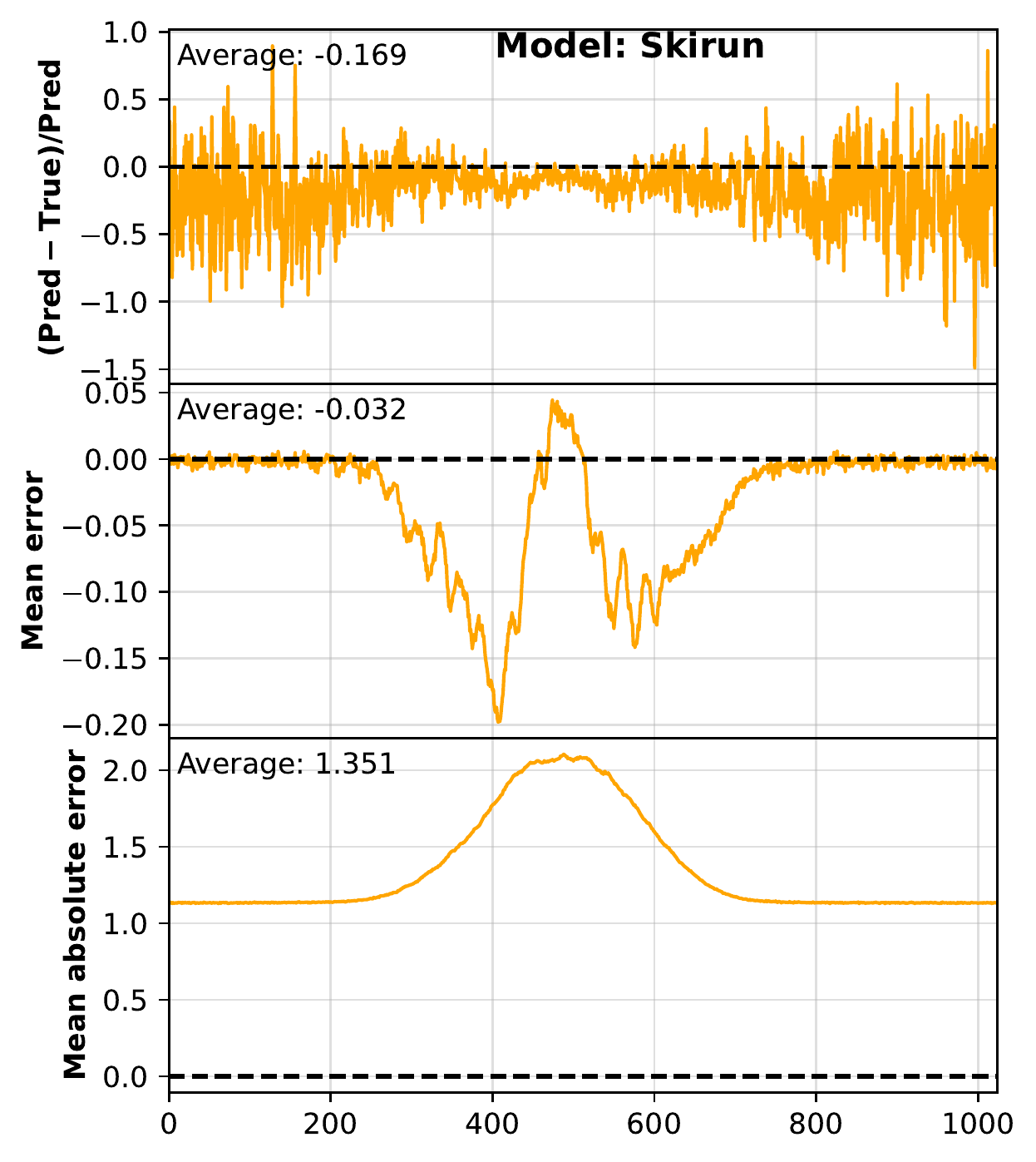}
  
  \includegraphics[width=0.32\textwidth]{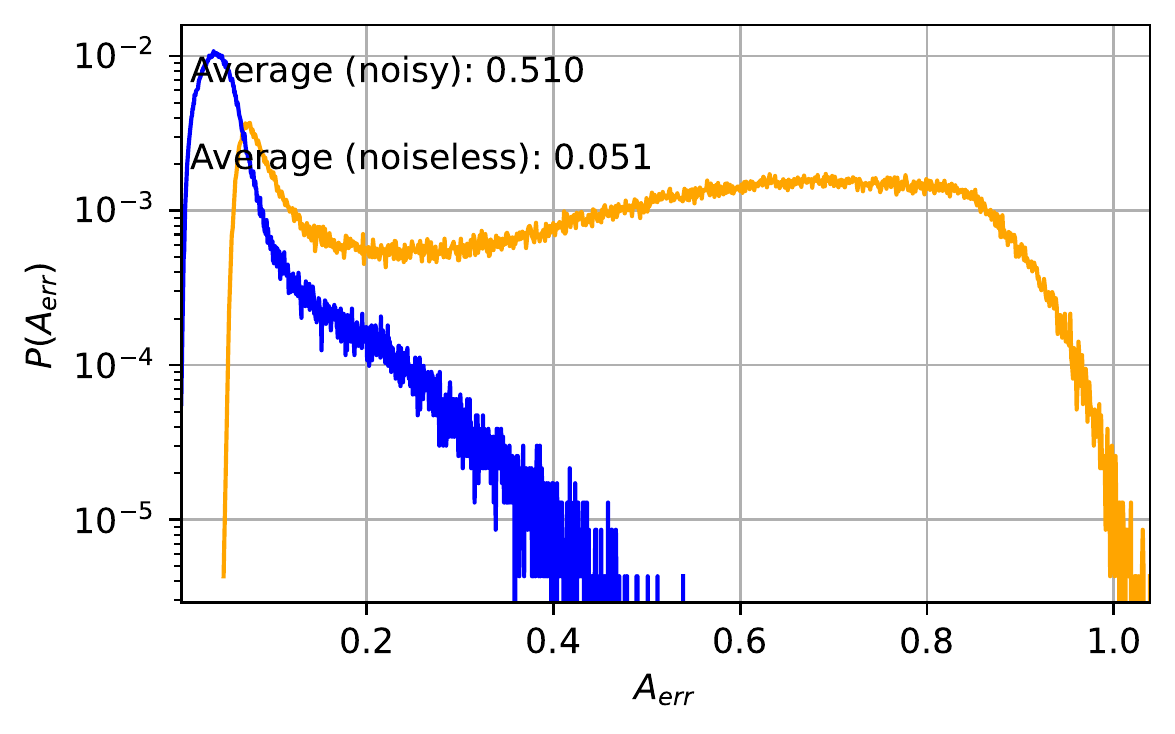}
  \includegraphics[width=0.32\textwidth]{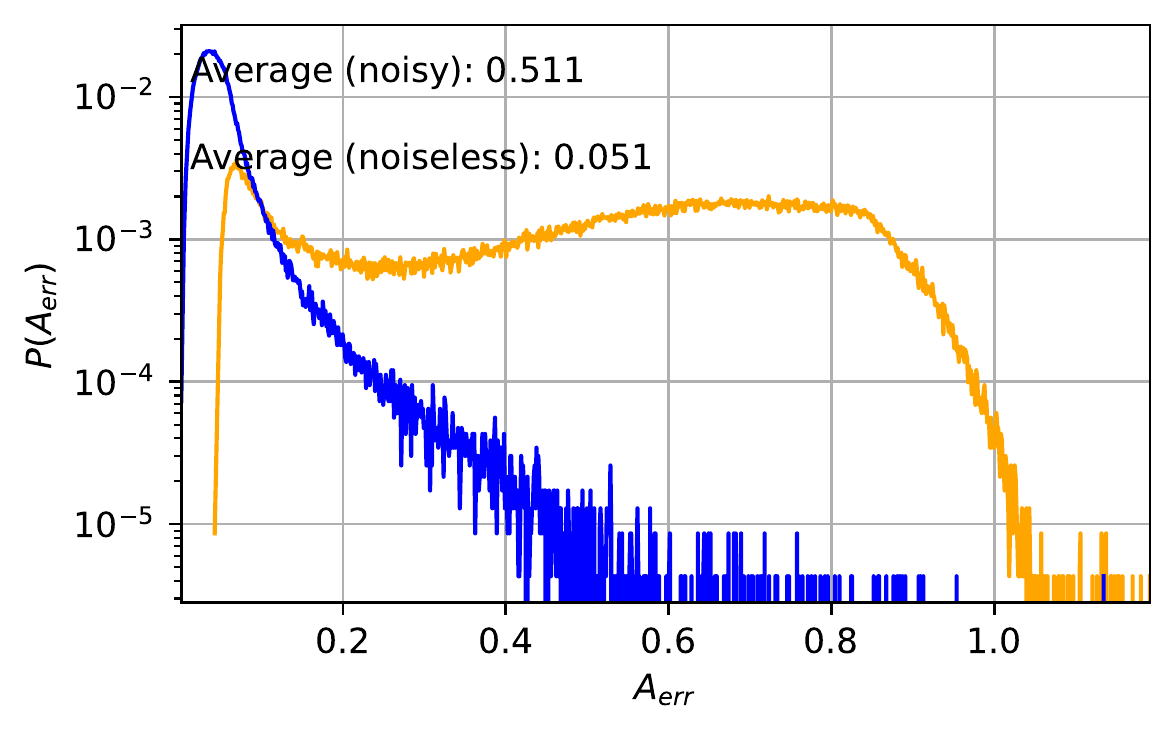}
  \includegraphics[width=0.32\textwidth]{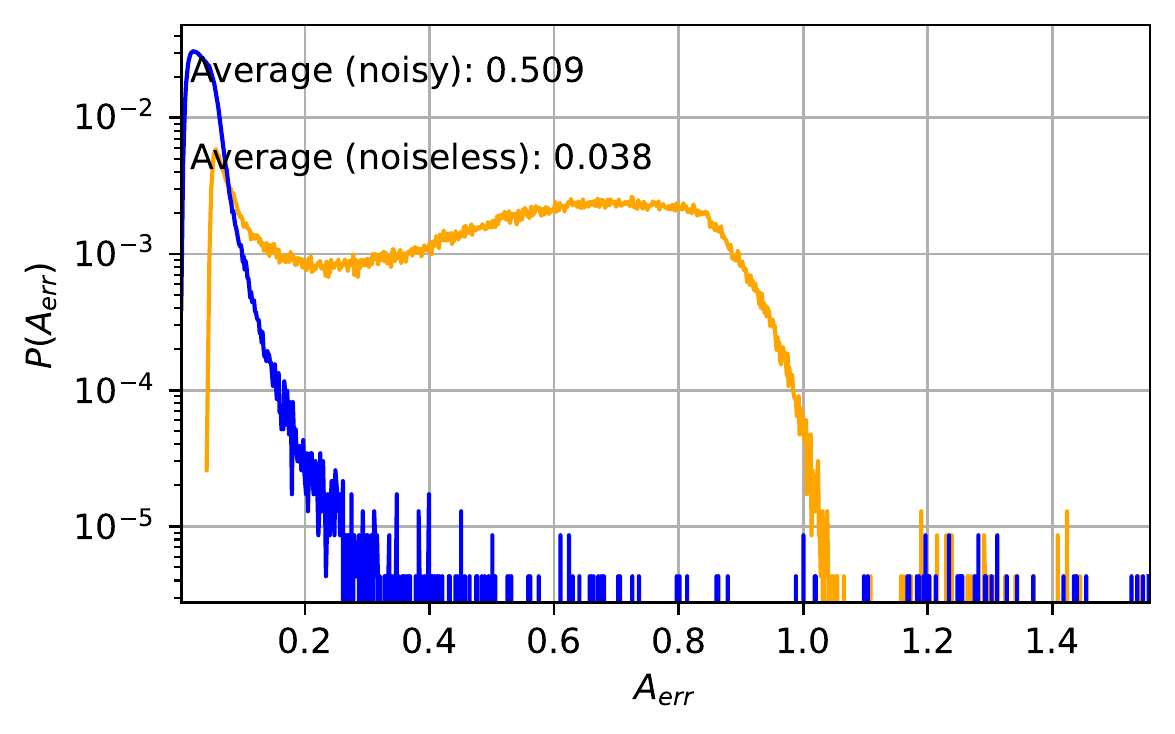}
  
  \includegraphics[width=0.32\textwidth]{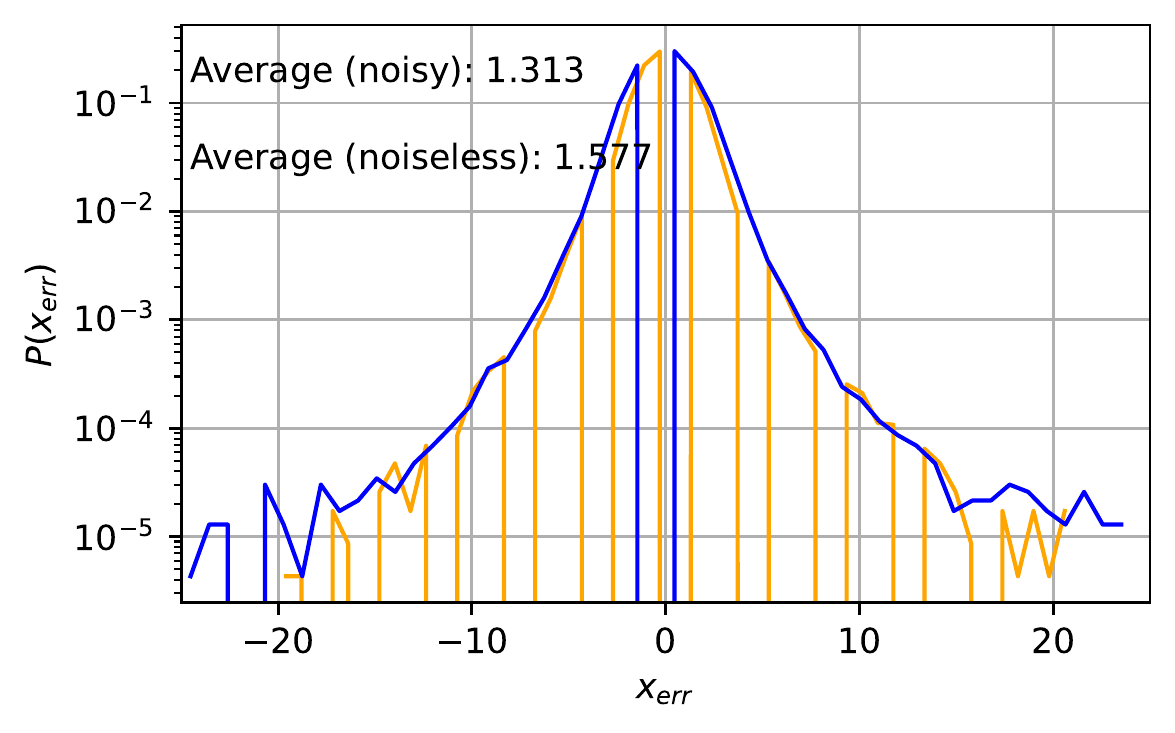}
  \includegraphics[width=0.32\textwidth]{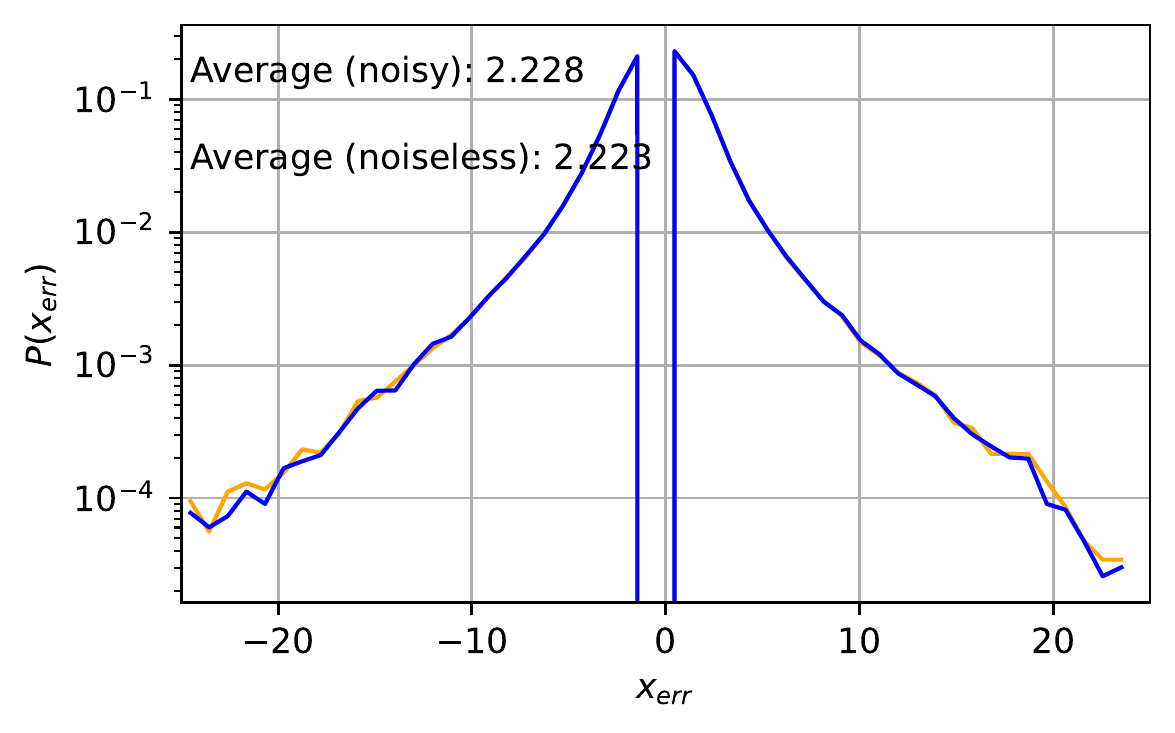}
  \includegraphics[width=0.32\textwidth]{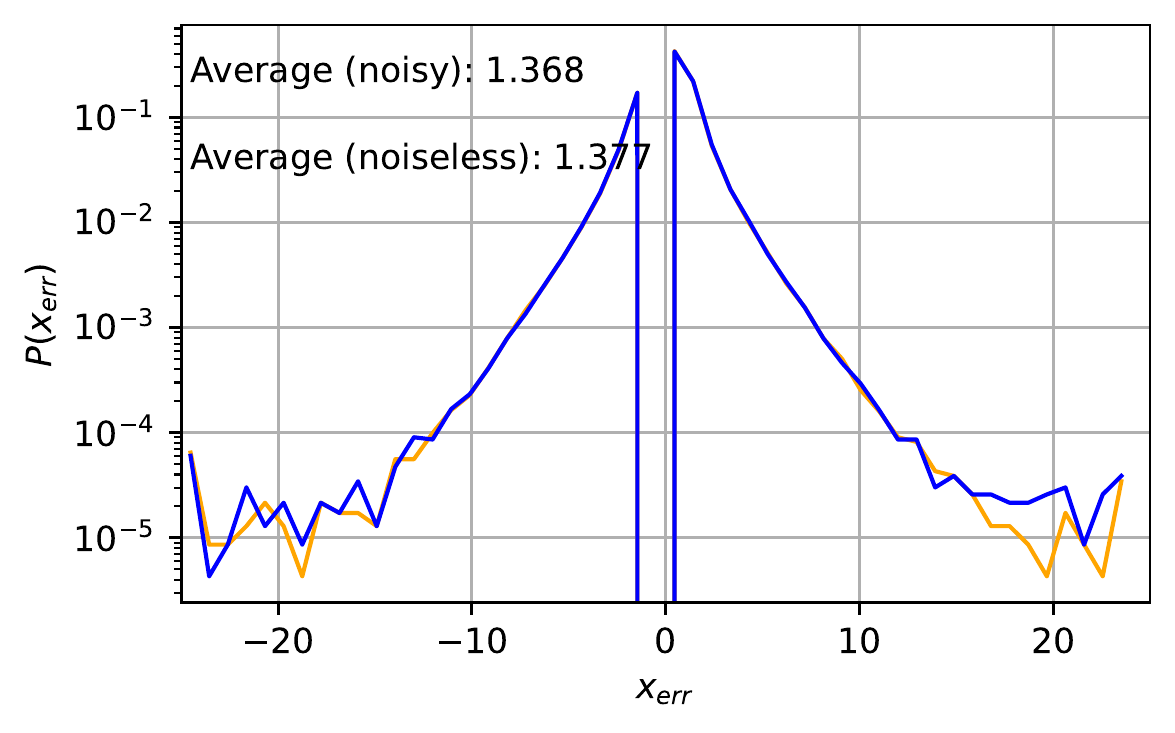}
  
  \caption{The average error values (see text for details).}
  \label{fig:avg}
\end{figure}

We show the probabilities of errors versus the errors  on the bottom two panels of Fig.~\ref{fig:avg}. The orange and blue curves correspond to comparing the predicted signal to the noisy and noiseless signals, respectively. As discussed above, both the probability histogram and the mean value of the amplitude error for the noiseless case shows the validity of the approximation given in Eq.~\eqref{eq:approx-accurate}. For every model, the relative error of the amplitude remains below $10 \, \%$ with a mean value of around $5 \, \%$. The severe distortion of the noise can also be seen on the figures. The long tail of the noisy histograms suggests that the vast majority of the samples are weighted with a significant amount of noise. However, considering the phase errors, we cannot see any difference between the noisy and the noiseless probability histograms as they overlap. This suggests that the predictions carry a very small amount of phase error. More precisely said, most prediction does not suffer a phase error at all. For all the models, only an unnoticeable fraction of the predictions carries a phase error greater than $10$ pixels. The mean value of the absolute phase errors reveals that the vast majority of the predictions has a phase error less than $2$ pixels, corresponding to a relative phase error of $0.1 - 0.2 \, \%$.


\FloatBarrier

Finally, Figure \ref{fig:examples} shows some examples for the predicted parameters and the Schlieren signals for each network architectures.

\subsection{Robustness of the models}
\label{subsec:robusttest}


Whether trained neural networks may be useful for the evaluation of real measurements also depends on the sensitivity of their results on parameter changes of the experiment. The networks have been trained on samples calculated with a set of fixed parameters, but some parameters can actually fluctuate or drift somewhat during the course of a measurement. This means that the neural networks will be used for evaluating samples that have been obtained with slightly different experimental parameters, therefore it is important to verify that network predictions are not affected substantially by such slight changes. In the setup considered here, the vapor density and the probe laser intensity are the most important parameters that may change slightly. The vapor density can be held constant to better than $1 \%$ accuracy, while the probe laser power may change possibly by a few per cents. In order to evaluate the effect of parameter changes, we have generated sets of test samples with vapor density decreased/increased by $2\%$, $5\%$ and $10\%$, respectively, as well as sets where the probe laser power was changed by the same amount. The mean absolute error $\sigma$ of parameter prediction was calculated in each case, as well as the mean error $\delta$, whose increase signifies systematic errors of the prediction. Please note that the purpose of this section is to test the robustness of the trained models---however, these changes in the vapor density and probe laser intensity could be easily incorporated in additional training datasets, that could further improve the neural networks' ability to 
give accurate and reliable predictions to the plasma parameters.

\begin{figure}[htb]
\includegraphics[width=0.8\textwidth]{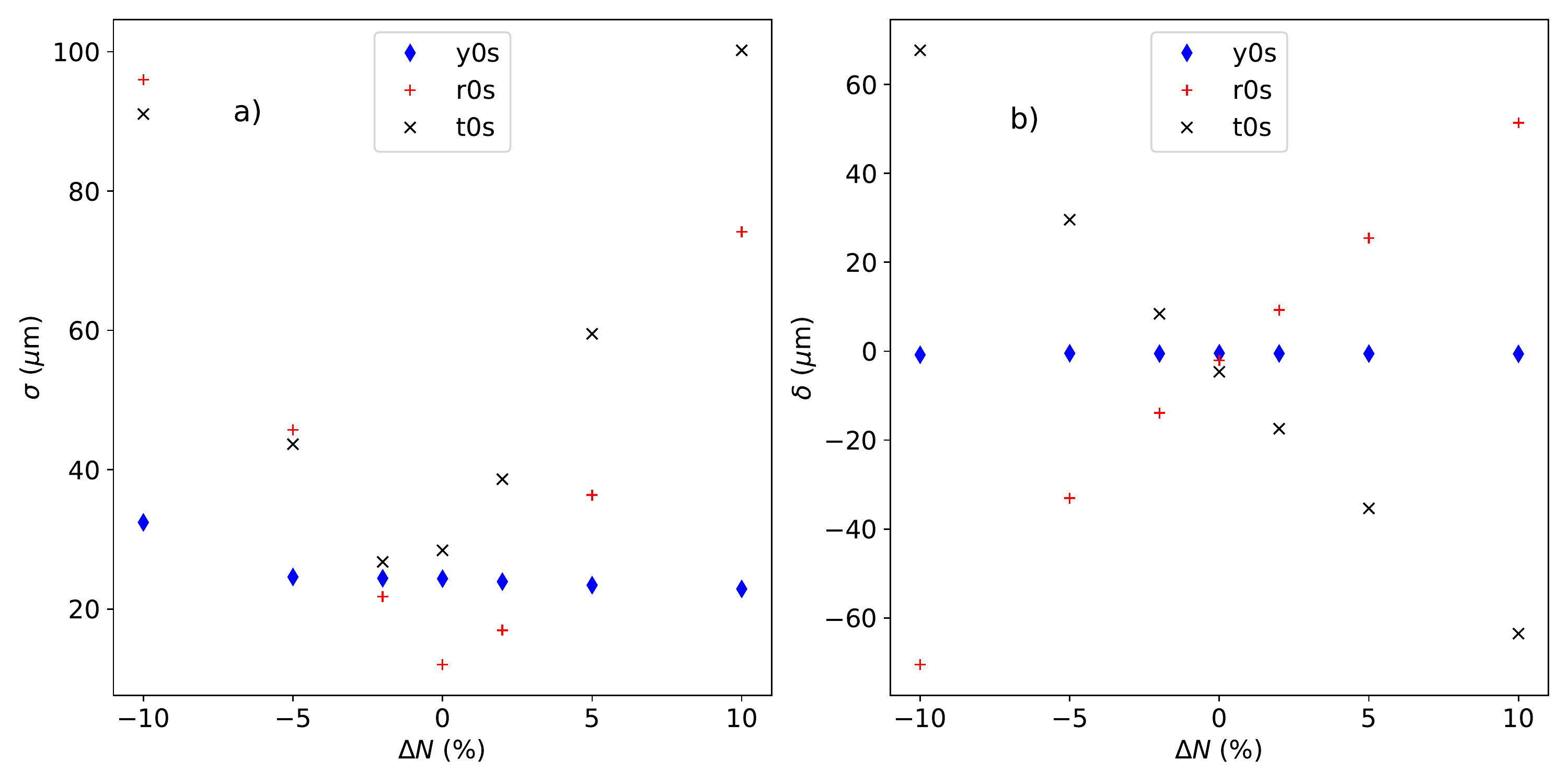}
\caption{a) Mean squared error $\sigma$ and b) mean error $\delta$ of parameter prediction as a function of vapor density change from the value used for training.  }
\label{density_variation}       
\end{figure}

\begin{figure}[htb]
\includegraphics[width=0.8\textwidth]{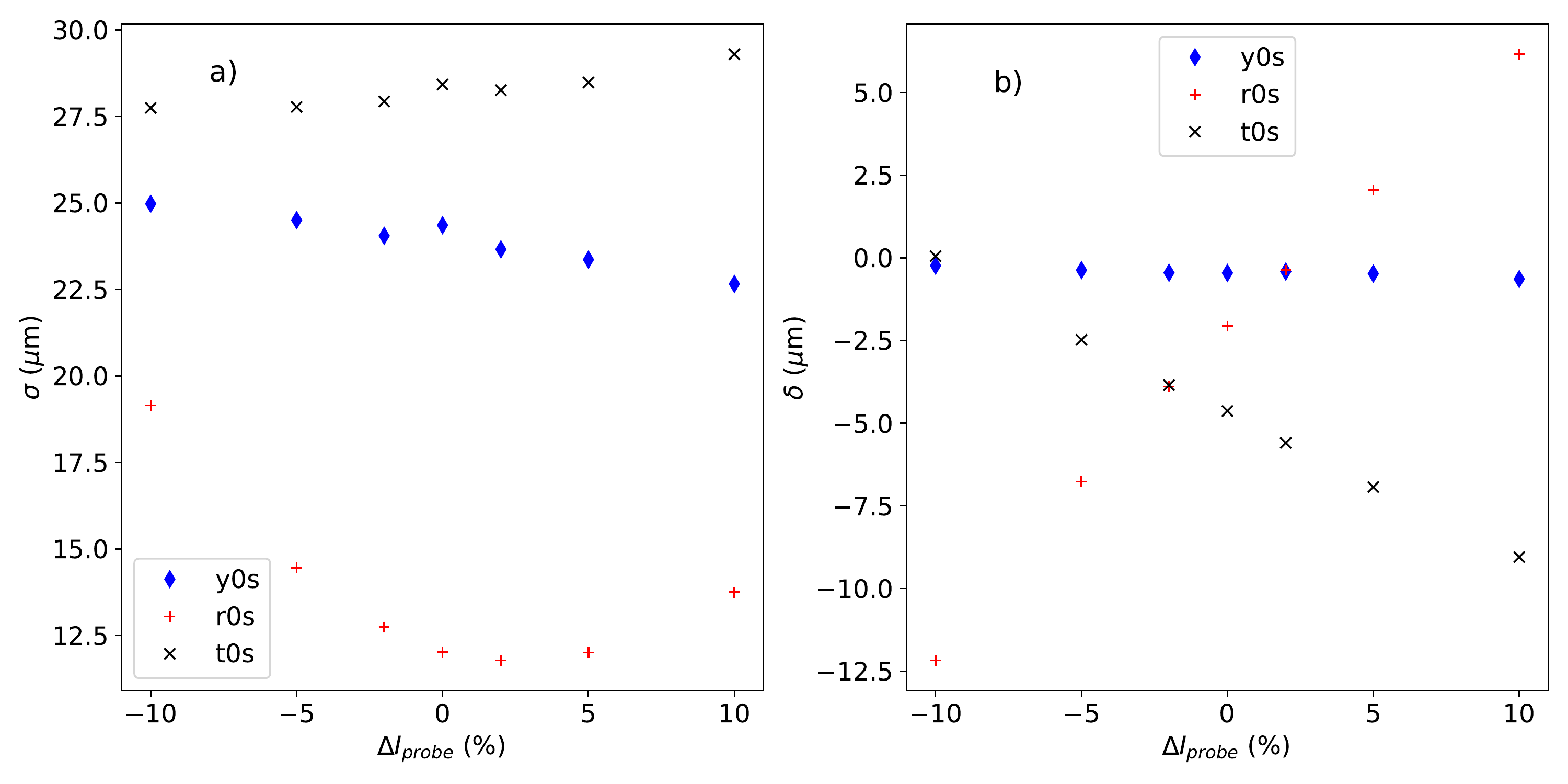}
\caption{a) Mean squared error $\sigma$ and b) mean error $\delta$ of parameter prediction as a function of probe laser power change from the value used for training. }
\label{probe_variation}       
\end{figure}

Figure \ref{density_variation} a) shows the increment of parameter prediction error as the vapor density deviates from the value used during training. The mean error $\delta$ for the same parameter predictions can be seen on Fig. \ref{probe_variation}. (Note that the values are plotted in $\mu$m.) It is interesting, that while $\sigma$ does not increase substantially for a deviation of 2\% for any of the three parameters plotted, $\delta$ does deviate from the ideal value for $r_0$ and $t_0$ in opposite sense. Nevertheless, a systematic error $<20\mu$m for $\Delta\mathcal{N}=2$\% is completely negligible. Probe laser power changes have a much smaller effect (see Fig. \ref{probe_variation}), even for a power variance of $\pm 10$\%.

\section{Summary}

In this paper a novel method to predict the geometrical dimensions of a plasma channel in atomic vapor has been presented, applying machine learning techniques. Schlieren imaging is an important method used to monitor the plasma properties, which is a crucial task in plasma wakefield acceleration technology. The implemented DNNs provide a robust and efficient framework to predict the plasma parameters from the noisy Schlieren signals with high accuracy. Without putting any constraint on the networks, they recognized the physical ranges of the parameters describing the plasma channel. The robustness of the networks has also been presented with respect to slight changes of the power of the probe laser and the density of the vapor. Improving this robustness may be achieved by further training the presented models on datasets corresponding to different powers of the probe laser and different vapor densities. These results and considerations suggests that after some improvement, the presented networks will be eligible to reliably evaluate the corresponding experimental data.

\section{Acknowledgment}

The research was supported by the Hungarian National Research, Development and Innovation Office (NKFIH) under the contract numbers OTKA K135515, K123815 and NKFIH 2019-2.1.11-T\'ET-2019-00078, 2019-2.1.11-T\'ET-2019-00050, 2019-2.1.6-NEMZ\_KI-2019-00011, 2020-2.1.1-ED-2021-00179, and the Wigner Scientific Computational Laboratory (WSCLAB, the former Wigner GPU Laboratory). Author G. B. was supported by the Ministry of Innovation and Technology NRDI Office within the framework of the MILAB Artificial Intelligence National Laboratory Program. The authors are grateful to Gábor Papp and Gergely Gábor Barnaföldi for the useful discussions.

\clearpage


\section*{References}

\input{schlieren1.bbl}

\end{document}

%% file: mathcommands.tex
\newcommand{\parenth}[1]{\left( #1 \right)}
\newcommand{\bparenth}[1]{\left[ #1 \right]}

\newcommand{\abs}[1]{\left\lvert #1 \right\rvert}

\newcommand{\avg}[1]{\left\langle #1 \right\rangle}

\newcommand{\diff}{\mathrm{d}}

